\documentclass[aps,pra,amsmath,amssymb,twocolumn,superscriptaddress,showpacs,10pt]{revtex4-1}

\usepackage{amsmath,amssymb,bbm,graphicx,color}

\pdfoutput=1

\newcommand{\be}{\begin{equation}}
\newcommand{\ee}{\end{equation}}
\newcommand{\ba}{\begin{array}}
\newcommand{\ea}{\end{array}}
\newcommand{\bqa}{\begin{eqnarray}}
\newcommand{\eqa}{\end{eqnarray}}

\newcommand{\tr}{\mbox{Tr}}
\newcommand{\di}{\text{d}}
\newcommand{\bra}[1]{\ensuremath{\langle #1 |}}
\newcommand{\ket}[1]{\ensuremath{| #1 \rangle}}

\newcommand{\ovl}[2]{\ensuremath{\langle #1 | #2 \rangle}}

\newcommand{\ie}{{\it i.e.\ }}
\newcommand{\eg}{{\it e.g.\ }}

\begin{document}

\title{Tailoring many-body entanglement through local control}

\author{Felix Lucas}
\affiliation{Max Planck Institute for the Physics of Complex Systems, N\" othnitzer Str. 38, D-01187 Dresden, Germany}
\affiliation{Physikalisches Institut, Albert-Ludwigs-Universit\"at Freiburg, Hermann-Herder-Strasse 3, D-79104 Freiburg, Germany}

\author{Florian Mintert}
\affiliation{Freiburg Insitute for Advanced Studies, Albert-Ludwigs-Universit\"at Freiburg, Albertstr. 19, D-79104 Freiburg, Germany}
\affiliation{Physikalisches Institut, Albert-Ludwigs-Universit\"at Freiburg, Hermann-Herder-Strasse 3, D-79104 Freiburg, Germany}

\author{Andreas Buchleitner}
\affiliation{Physikalisches Institut, Albert-Ludwigs-Universit\"at Freiburg, Hermann-Herder-Strasse 3, D-79104 Freiburg, Germany}

\date{\today}

\begin{abstract}
We construct optimal time-local control pulses based on a multipartite entanglement measure as target functional. The underlying control Hamiltonians are derived in a purely algebraic fashion, and the resulting pulses drive a composite quantum system rapidly into that highly entangled state which can be created most efficiently for a given interaction mechanism, and which bears entanglement that is robust against decoherence. Moreover, it is shown that the control scheme is insensitive to experimental imperfections in first order.
 \end{abstract}
 
\maketitle 

 When we think of entanglement or quantum correlations as an elusive quantum phenomenon, we generally bear two empirical aspects of entanglement in mind: either it is shared between only very few particles, or it decays so rapidly that it is hard to observe experimentally. This dilemma is due to the fact that correlations between individual quantum objects can only arise through interactions. But particles that interact strongly with each other, such as ions or spins in a solid, typically also couple to their environment rather strongly, so that decoherence degrades the entanglement that is being created through interactions \cite{PhysRevLett.106.130506,Neumann06062008}. On the other hand, a weakly or non-interacting particle, such as a photon, is very unlikely to get entangled with multiple partners at the same time \cite{Yao2012}, but does not suffer from rapid decay of entanglement~\cite{ISI:000248219100018}.

 Therefore, if we aim to create many-body entanglement that persists over long periods of time, it is imperative to deal with the interplay between interactions and decoherence. Our plan in the present paper is to apply the principles of quantum control in order to optimally exploit a system's intrinsic interactions and, at the same time, to reduce the detrimental influence of decoherence.

 More explicitly, a system that naturally evolves under a system Hamiltonian $H_\text{sys}$ and couples to its environment can be steered with an additional time-dependent control Hamiltonian $H_\text{c}(t)$, induced, for example, by shaped laser pulses. For weak environment coupling, the system dynamics is described by a master equation in Lindblad form \cite{lindblad}
 \be
  \dot \rho = - \frac i \hbar [H_\text{sys} + H_\text{c},\rho] + \underbrace{\sum_i  \gamma_i \left( L_i \rho L_i^\dagger - \frac 1 2 \left\{L_i^\dagger L_i, \rho\right\} \right)}_{{\cal D}(\rho)},
  \label{eq:drho}
 \ee
 with an incoherent part ${\cal D}(\rho)$ that is due to the environment.
 Since we aim at creating entanglement, the target of control will be formulated here in terms of entanglement measures, \ie its most natural quantifier. A time-local quantum control strategy~\cite{alessandrocontrol} for our target then enables us to find that $H_\text{c}(t)$, for which entanglement is maximized under the dynamical constraints set by Eq.~\eqref{eq:drho}. Not surprisingly, the solution of this control problem is optimal if the created entanglement is optimally adapted to the dynamics \eqref{eq:drho}: \ie if it is both robust under decoherence, and efficiently creatable through the system interactions \cite{PRL}.

 We will hence proceed by first defining our target as a functional $\tau(\rho)$ of the system state in section~\ref{sec:targetfunc}. Subsequently, in section~\ref{sec:strategy}, we outline the time-local strategy to achieve the maximization of $\tau$, and algebraically derive the corresponding optimal time-local control Hamiltonian. In section~\ref{sec:singlecontrol}, this strategy is specialized to the situation of restricted, single-particle control fields and subsequently numerically applied to an exemplary physical system which exhibits both interactions and environment coupling (sections~\ref{sec:cohcontrol} and \ref{sec:dissipativesystems}). Finally, in sections~\ref{sec:robuststates} and \ref{sec:pulserobustness} we will elaborate on the intrinsic robustness of our control strategy with respect to experimental imperfections and decoherence.
  
\section{The Target of Control}
\label{sec:targetfunc}

 In general, quantum control can be formulated in terms of a functional $\tau(\rho)$ of the system state, called \emph{objective} or \emph{target functional}, that is maximized by application of suitably tailored control fields. The most frequently applied target functional is fidelity $\bra{\psi_\text{tar}}\rho\ket{\psi_\text{tar}}$
  \cite{tannor:5013,PhysRevA.37.4950,glaserGRAPE}, which quantifies the similarity of the system state to a specific, predefined target state $\ket{\psi_\text{tar}}$. If we choose a highly entangled multipartite state as target state, then the achievement of maximal fidelity implies the success of creating entanglement in the system.

 There exists, however, a multitude of states with equal entanglement properties---those, that are connected via single-particle (local) unitary transformations \cite{PhysRevLett.104.020504}---each of which may display distinct dynamical behavior. Among these states, some will show a more rapid decay of entanglement under decoherence than the average, whereas some will feature rather robust entanglement \cite{flodeco,PhysRevA.65.052327}. Similarly, some might be very hard to produce under the given intrinsic system interactions, whereas others are better suited for a specific interaction mechanism. To identify the most suitable entangled states---\eg those which are robust under a given environmental coupling, or those which are easily and rapidly produced---is a highly non-trivial question, which has only recently moved into the focus of research~\cite{borplasrob,florobust}.
 
 By definition, entanglement measures \emph{directly} quantify the entanglement of a system, irrespective of the specific state the entanglement is inscribed in---in other words, they are invariant under local unitary transformations. Therefore, the choice of an entanglement measure as target functional allows to maximize entanglement without the unnecessary and possibly disadvantageous restriction to one specific target state \cite{PRL}. What is more, as we will see shortly, with this choice for the target functional the most suitable system states mentioned above are naturally created in the control process.
 
 Although the optimization of two-qubit entanglement based on measures such as concurrence \cite{woocon} was achieved before for pure states \cite{entgates,loccon}, an extension to mixed states in many-body systems requires efficiently accessible entanglement measures, and therefore was out of reach for a long time. Typically, entanglement measures $E(\rho)$ for \emph{mixed} states are defined via an optimization problem, for example in terms of a given \emph{pure} state entanglement measure $E(\psi)$. An optimization is necessary to make the connection between $E(\rho)$ and $E(\psi)$ since a mixed state $\rho$ can be represented by infinitely many pure state ensembles $\{p_i,\ket{\psi_i}\}$, with $\rho = \sum_i p_i \ket{\psi_i}\bra{\psi_i}$, which, in general, bear different average amounts of entanglement. A valid mixed state entanglement measure is given by the \emph{lowest} average pure state entanglement of all possible ensembles \cite{PhysRevA.62.032307}, \ie
 \be
 	E(\rho) = \inf_{\{p_i,\ket{\psi_i}\}} \sum_i p_i E(\psi_i)\ .
 	\label{eq:convexroof}
 \ee
 An explicit and exact expression for this infimum is only known for the frequently utilized measure concurrence for two qubits~\cite{woocon}. Generically, the optimization in Eq.~\eqref{eq:convexroof} can be performed at best for single states $\rho$, but becomes computationally intractable if to be evaluated over and over again for many different states, as is necessary for the use as a target functional.

 We will therefore resort to a \emph{lower bound} of $E(\rho)$ for a many-body generalization
 \bqa
 	E(\psi) = \sqrt{\bra \psi \otimes \bra \psi \mathbf{V} \ket \psi \otimes \ket \psi} \text{, with} 	\label{eq:genconc}\\
 	\mathbf{V} = 4 \sum P_\pm^1 \otimes \ldots \otimes P_\pm^N, \nonumber
 \eqa
of concurrence \cite{flopure}. The $P_\pm^i$ are projectors onto the symmetric and antisymmetric subspace of the duplicate Hilbert space $\mathcal H_i \otimes \mathcal H_i$ of subsystem $i$, and the sum runs over all combinations of $P_\pm^i$ that involve an even number of projectors onto antisymmetric subspaces. The corresponding lower bound for mixed states \cite{flomix, aolita:022308}
 \bqa
 	&\tau &= \tr(\rho \otimes \rho \, \mathbf{A}) \leq E(\rho)^2,\mbox{ with} \label{eq:mpconc2}\\
	&&\mathbf{A}={\bf V}-4(1-2^{1-N}) \boldsymbol{P}_{-} \nonumber,
 \eqa
 is defined in terms of the projector $\boldsymbol{P}_{-}$ onto the anti-sym\-metric component of the duplicate many-body Hilbert Space $\mathcal H \otimes \mathcal H$.
 
 The functional $\tau$ satisfies all desired properties without suffering from the prohibitive mathematical difficulties of an exact measure:
 \begin{itemize}
 \item[-]It is a purely algebraic function of the system's density matrix $\rho$ and hence is readily implemented as figure of merit in quantum control.
 \item[-]It is invariant under local unitary dynamics, thus treating all equally entangled states equally.
 \item[-]The lower bound's accuracy increases with increasing purity of the density matrix $\rho$. Since the most strongly entangled states are pure, a suitable control field will drive a system towards pure states, so that the approximation implicit in Eq.\eqref{eq:mpconc2} is getting exact as the optimization's goal is reached.
 \end{itemize}
 
\section{The Control Strategy}
 \label{sec:strategy}
 With a suitable target functional at hand, we can now strive for the optimization of entanglement dynamics by means of externally applied control fields. For that purpose, we will apply a time-local control strategy based on Lyapunov control \cite{Mirrahimi20051987,alessandrocontrol}, which achieves the maximization of the target functional by maximization of its \emph{time-derivatives}. The working principle of this strategy is illustrated by the fact that a maximized first derivative $\dot \tau$ will subsequently lead to a more rapidly increasing target functional $\tau$. This also holds for higher derivatives of $\tau$: \eg maximizing the curvature $\ddot \tau$ induces an increase of the first derivative $\dot \tau$, which, in turn, spurs the augmentation of $\tau$.
 
 
 The first time derivative of our target functional \eqref{eq:mpconc2} reads
 \be
 	\dot \tau = \tr(\dot \rho \otimes \rho \mathbf{A} + \rho \otimes \dot \rho \mathbf{A}) = 2 \tr(\dot \rho \otimes \rho \mathbf{A}),
  \label{eq:dtau}
 \ee
 where the second equality holds since the expectation value of $\mathbf A$ is invariant under permutation of $\rho$ and $\dot \rho$ \cite{aolita:022308}.

 By virtue of Eq.~\eqref{eq:drho}, $\dot \tau$ can be recast into the form
 \be
  \dot \tau = 2 \tr\left\{ \left( -\frac i \hbar \left[ H_\text{sys} +H_\text{c},\rho \right] + {\cal D}(\rho) \right) \otimes \rho \; \mathbf{A} \right\},
  \label{eq:dtau2}
 \ee
 and, since $\dot \tau$ depends linearly on $H_\text{c}$, we can describe the influence of external control by the partial derivatives of $\dot \tau$ with respect to the free parameters in the control Hamiltonian $H_\text{c}$
 \be
 	\dot \tau = \frac{\partial \tau} {\partial H_\text{c}} H_\text{c} + \dot \tau_0.
 	\label{eq:Hcderi}
 \ee
 Here the term $\dot \tau_0$ represents the natural dynamics of $\tau$ in the absence of control.
 
 To be more explicit, for $n$ finite-dimensional subsystems, we can expand $H_\text{c}$ in a suitable operator basis as
 \be
  H_\text{c} = \sum_{\{\xi_1,\ldots,\xi_n\}} \beta_{\xi_1}^{(1)} \otimes \ldots \otimes \beta_{\xi_n}^{(n)} \; h_{\vec \xi} = \vec \beta \cdot \vec h,
  \label{eq:generalcontrolhamiltonian}
 \ee
 where the superscript $(n)$ indicates that the basis element $\beta_{\xi_n}^{(n)}$ acts on the $n^\text{th}$ particle's Hilbert space, and $\xi_n$ labels the basis elements of the corresponding subspace. The expansion coefficients $h_{\vec \xi}$, which encode the strength of the different components of the control Hamiltonian, are appropriately labeled such that they represent a single vector~$\vec h$. In this notation, the partial derivative with respect to $H_\text{c}$ in Eq.~\eqref{eq:Hcderi} turns into the partial derivatives with respect to the components of $\vec h$:
 \be
  \dot \tau = \sum_i \frac{\partial \tau}{\partial h_i} h_i + \dot \tau_0 = \vec h \cdot \vec Y (\rho,H_\text{sys},{\cal D}) + \dot \tau_0.
  \label{eq:dtauscalprod}
 \ee
 In this form, one can see that $\dot \tau$ can be conveniently written as a scalar product of the vector $\vec Y$, encompassing the partial derivatives $\partial \tau /\partial h_i$, with $\vec h$. $\vec Y$ is an algebraic expression of the system's state $\rho(t)$, of the inter-particle interactions $H_\text{sys}$, and of the incoherent evolution $\cal D$ induced by environmental noise. Finally, from Eq.~\eqref{eq:dtauscalprod} we can read off what will be the optimal control Hamiltonian that maximizes the first derivative of our target functional: for a given maximal norm of the vector $\vec h$, we have to choose $\vec h \parallel \vec Y$ to maximize $\dot \tau$.
 
 Again, the nature of these control fields $\vec h$ is strictly time-local---\ie they result in a maximal time derivative $\dot \tau(t)$ in one particular instant $t$, and depend on the system state at that time alone, $\vec h(\rho(t))$. In contrast, an optimization over the whole control time interval which is performed in \emph{optimal control} \cite{tannorkrotov,rabitzkrotov,glaserGRAPE}, yields globally optimal pulses with potentially better performance. However, not only that our straightforward time-local strategy is numerically significantly less demanding, since it requires only one single propagation of the control system (optimal control algorithms such as KROTOV \cite{rabitzkrotov} and GRAPE \cite{glaserGRAPE} require many iterations of forward and backward propagation). It has also proven to create maximum entanglement on the same time scales as the aforementioned time-global control strategies, as we will see in the numerical studies in Sec.~\ref{sec:cohcontrol}.

\section{Restricted Control: Manipulating Single Particle Dynamics}
\label{sec:singlecontrol}
 
 In the preceding section we have derived control fields that maximize $\dot \tau$ under the assumption that an arbitrary control Hamiltonian $H_\text{c}$ such as \eqref{eq:generalcontrolhamiltonian} is at our disposal. However, a control scheme designed for application to a broad range of physical systems has to take into account that experimentally implementable controls may be restricted---\eg in many experiments it is difficult or even virtually impossible to engineer inter-particle interactions. This means that, realistically, we have to assume that $H_\text{c}$ can induce only local unitary time evolution of $\rho$, \ie a \emph{local} control Hamiltonian (not to be confused with the meaning of a time-local control strategy explained above). In the present section we will derive the control Hamiltonian under this restriction.
 
 The most general local Hamiltonian for $n$ finite dimensional particles reads
 \be
 	H_\text{c} = \sum_{i=1}^n \sum_{\xi_i = 1}^{d_i}  h_{\xi_i}^{(i)} \, \beta_{\xi_i}^{(i)},
	\label{eq:localhc}
 \ee
 where the operators $\beta_{\xi_i}^{(i)}$ form a basis of the $i^\text{th}$ particle's Hilbert space.
For ease of notation, in the following we will consider two-level systems (\eg spin-$1/2$ particles) and choose the Pauli matrices $\beta_{\xi_i}^{(i)} = \sigma_\xi^{(i)}$ ($\xi \in \{ x, y, z\}$) as operator basis, although our derivation remains valid also for higher dimensional systems.
 
 Here, it is important to recall that our control functional~\eqref{eq:mpconc2} is invariant under the local unitary dynamics induced by a local Hamiltonian. Therefore, $\dot \tau$ will not depend on $H_\text{c}$. In general, only non-local Hamiltonians (\eg interactions) or local, but non-unitary, dynamics (\eg decoherence) can induce a change of entanglement.
 
 Since $\dot \tau$ does not reveal the influence of {\em local} control on the entanglement dynamics, we have to resort to $\ddot\tau$ to identify the optimal control Hamiltonian.
In contrast to $\dot \tau$, which describes the entanglement dynamics in first order in d$t$, $\ddot \tau$ describes the change of entanglement in two infinitesimal time steps. $\ddot\tau$ thus contains the information of how the impact of interactions and decoherence in a second time-step is affected if the system is manipulated locally in the first step. Formally, this is reflected in the bilinear dependence of $\ddot\tau$ on the generator of the system dynamics, that gives rise to cross-terms of $H_\text{c}$ and $H_\text{sys}$ or ${\cal D}$. As we will see shortly, this \emph{interplay} of local, coherent control, on the one hand, and of system interactions or incoherent dynamics, on the other, enables us to control the entanglement dynamics through local manipulation only.
 
 Let us spell this out, for the sake of clarity. Our objective is now the maximization of the curvature
 \be
 	\ddot \tau = 2 \tr\Big(\{\ddot \rho \otimes \rho + \dot \rho \otimes \dot \rho\} \mathbf A\Big).
	\label{eq:ddtau}
 \ee
 This curvature $\ddot \tau$ is a function of $\dot \rho$ and $\ddot \rho$, given by the equation of motion \eqref{eq:drho}, and by the time derivative of Eq.~\eqref{eq:drho}, respectively. Since only the state $\rho$ and the control Hamiltonian $H_\text{c}$ in Eq.~\eqref{eq:drho} are time-dependent, we have
 \be
 	\ddot \rho = - \frac i \hbar \left([\dot H_\text{c},\rho] + [H_\text{sys} + H_\text{c}, \dot \rho] \right) + \mathcal{D}(\dot \rho),
 \ee
 which can be specified in terms of $\rho$ by iteratively substituting $\dot \rho$ with the equation of motion~\eqref{eq:drho}:
 \be
 \begin{split}
 	\ddot \rho &= - \frac 1 {\hbar^2} \big[ H_\text{sys} +\! H_\text{c},[H_\text{sys} +\! H_\text{c}, \rho] \big] - \frac i \hbar [H_\text{sys} +\! H_\text{c}, {\cal D}(\rho)] \\
	& \mathrel{\phantom{=}} - \frac i \hbar {\cal D} \big([H_\text{sys} + H_\text{c}, \rho] \big) + {\cal D}({\cal D}(\rho)) - \frac i \hbar [\dot H_\text{c}, \rho].
	\label{eq:ddrho}
 \end{split}
 \ee
 As expected, $\ddot \rho$ involves bilinear terms in the Hamiltonian $H_\text{sys} + H_\text{c}$ and the Lindbladian $\mathcal D$.

 Finally, an expression for the curvature of $\tau$ with explicit dependence on the control Hamiltonian is obtained by employing $\dot\rho$ and $\ddot\rho$ given by Eqs.~\eqref{eq:drho} and \eqref{eq:ddrho} in Eq.~\eqref{eq:ddtau}:
\be
 \begin{split}
 	\ddot \tau &= \tr \left\{\left( - \frac 1{\hbar^2}\Big([H_\text c, [H_\text c,\rho]] + [H_\text{sys}, [H_\text c,\rho]] \right. \right. \\
	&  \mathrel{\phantom{=}} + [H_\text{c}, [H_\text{sys},\rho]] \Big) - \frac i \hbar \Big( [\dot H_\text{c}, \rho] + [H_\text c,\mathcal D(\rho)] \\
	&  \mathrel{\phantom{=}} + \mathcal D([H_\text c,\rho])\Big)\bigg) \otimes \rho -\frac 1 {\hbar^2} \Big( [H_\text{c},\rho] \otimes [H_\text{c},\rho] \\
	&  \mathrel{\phantom{=}} + [H_\text{sys},\rho] \otimes [H_\text{c},\rho] + [H_\text{c},\rho] \otimes [H_\text{sys},\rho] \Big) \\
	&  \mathrel{\phantom{=}} -\frac i \hbar \Big([H_\text{c}, \rho] \otimes \mathcal D(\rho) + \mathcal D(\rho) \otimes [H_\text{c}, \rho] \Big) \bigg\} \, \mathbf A + \ddot \tau_0,
 \end{split}
 \label{eq:ddtau2}
 \ee
 where terms that depend on $H_\text{c}$ have been separated from the natural curvature, \ie the curvature in the absence of control
 \be
 \begin{split}
 	\ddot \tau_0 &= \tr \left\{ \left( - \frac 1 {\hbar^2}[H_\text{sys},[H_\text{sys},\rho]] - \frac i \hbar [H_\text{sys},\mathcal D(\rho)] \right. \right.\\
	& \mathrel{\phantom{=}} - \frac i \hbar \mathcal D([H_\text{sys},\rho]) + \mathcal D(\mathcal D(\rho)) \bigg) \otimes \rho \\
	& \mathrel{\phantom{=}} - \frac 1 {\hbar^2} [H_\text{sys},\rho] \otimes [H_\text{sys},\rho] - \frac i \hbar [H_\text{sys},\rho] \otimes \mathcal D(\rho) \\
	& \mathrel{\phantom{=}}  - \frac i \hbar \mathcal D(\rho) \otimes [H_\text{sys},\rho] 	+ \mathcal D(\rho) \otimes \mathcal D(\rho) \bigg\} \, \mathbf A.
 \end{split}
 \label{eq:ddtau0}
 \ee
 Since $H_c$ induces only local unitary dynamics, the term in Eq.~\eqref{eq:ddtau2} which involves $H_\text{c}$ bilinearly, and the term that involves the time derivative $\dot H_\text{c}$ necessarily vanish. This can be seen by considering a purely coherent evolution of non-interacting particles---\ie $H_\text{sys} = 0$ and $\mathcal L = 0$. In this case, the system follows a purely local unitary time evolution under $H_\text{c}$, which, as argued above, leaves $\tau$ unchanged. Thus all derivatives of $\tau$ vanish, and in particular
 \be
 \begin{split}
 	\ddot \tau &= \tr \left\{ - \frac 1{\hbar^2}\Big([H_\text c, [H_\text c,\rho]]\otimes \rho + [H_\text{c},\rho] \otimes [H_\text{c},\rho] \Big) \right. \\
	&\phantom{\mathrel{=} \tr \bigg\{} \left. - \frac i \hbar [\dot H_\text{c}, \rho] \otimes \rho \right\} \, \mathbf A = 0.
 \end{split}
 \ee
 
 Eq.~\eqref{eq:ddtau2} for $\ddot \tau$ then reduces to those terms that contain both $H_\text{c}$ and $H_\text{sys}$, which represent the aforementioned interplay between the system interactions with external control, plus those terms that contain both $H_\text{c}$ and $\cal D$, \ie the interplay between control and decoherence. The former is of \emph{nonlocal} type, whereas the latter is \emph{non-unitary} and, hence, this interplay \emph{does} affect the system entanglement. It therefore allows us to change the dynamics of entanglement	 through local control.
 
 To put it in more physical terms: although local control itself cannot change the system entanglement, it can very well induce a change towards a state with different dynamical properties. In this sense, the interplay of $H_\text{c}$ and $H_\text{sys}$ reveals which local rotations {one needs} to apply, such that subsequently the intrinsic system interactions can efficiently entangle the constituent particles. More precisely, by maximization of the corresponding terms in Eq.~\eqref{eq:ddtau2}, we drive the system towards those states for which interactions create entanglement most rapidly. Analogously, the interplay of $H_\text{c}$ and $\cal D$ provides information on how to change the system state locally, to render its entanglement least vulnerable to the environmental influence, and the optimum is reached by maximization of the corresponding terms.
 In systems subject to both, interactions and decoherence, the maximum of the curvature $\ddot \tau$ \eqref{eq:ddtau2} typically corresponds to a trade-off between high robustness and high entangling efficiency. Therefore, the optimal states reached  \emph{automatically} by our control strategy are those highly entangled states which are rapidly produced and which are, at the same time, robust against environmental decoherence. Both properties are crucial when it comes to the experimental implementation of control, and we will investigate them in detail in Sec.~\ref{sec:cohcontrol} and \ref{sec:robuststates}.
 
 To find out what will be the \emph{optimal} control Hamiltonian that achieves the above objectives, we follow the same scheme as in Sec.~\ref{sec:strategy}. The curvature $\ddot \tau$ in Eq.~\eqref{eq:ddtau2} depends linearly on $H_\text{c}$, since, as argued above, bilinear terms vanish. Using the expansion \eqref{eq:localhc} of $H_\text{c}$, we can rewrite $\ddot \tau$ as
 \be
 	\ddot \tau = \sum_{i,\xi} \frac{\partial \ddot \tau}{\partial h_\xi^{(i)}} h_\xi^{(i)} + \ddot \tau_0 = \sum_i \vec h^{(i)} \cdot \vec X^{(i)} (\rho,H_\text{sys},{\cal D}) + \ddot \tau_0,
  \label{eq:ddtauscalprod}
 \ee
 with $\vec X^{(i)} = \sum_\xi \partial \ddot \tau / \partial h_\xi^{(i)}$. We obtain the partial derivatives from Eq.~\eqref{eq:ddtau2} with the help of Eq.~\eqref{eq:localhc} 
 \be
  \begin{split}
 	\frac{\partial \ddot \tau}{\partial h_\xi^{(i)}} &= \tr \left\{\left( - \frac 1{\hbar^2}\Big([H_\text{sys}, [\sigma_\xi^{(i)},\rho]] + [\sigma_\xi^{(i)}, [H_\text{sys},\rho]] \Big)\right. \right. \\
	&  \mathrel{\phantom{=}} \left.  - \frac i \hbar \Big([\sigma_\xi^{(i)},\mathcal D(\rho)] + \mathcal D([\sigma_\xi^{(i)},\rho])\Big)\right) \otimes \rho  \\
	&  \mathrel{\phantom{=}} -\frac 1 {\hbar^2} \Big( [H_\text{sys},\rho] \otimes [\sigma_\xi^{(i)},\rho] + [\sigma_\xi^{(i)},\rho] \otimes [H_\text{sys},\rho] \Big) \\
	&  \mathrel{\phantom{=}} \left. -\frac i \hbar \Big([\sigma_\xi^{(i)}, \rho] \otimes \mathcal D(\rho) + \mathcal D(\rho) \otimes [\sigma_\xi^{(i)}, \rho] \Big) \right\} \mathbf A,
 \end{split}
 \label{eq:partialderi}
 \ee
 and can thus compute the vectors $\vec X^{(i)}$. Since the expansion coefficients $h_\xi^{(i)}$ of the control Hamiltonian represent the amplitude of the three control fields ($\xi = x,y,z$) applied to the $i^\text{th}$ particle, the total amplitude directed at this particle is given by
 \be
 	\lVert \vec h^{(i)} \rVert = \sqrt{\big|h_x^{(i)}\big|^2 + \big|h_y^{(i)}\big|^2 + \big|h_z^{(i)}\big|^2}.
 \ee
 As before, we choose $\vec h^{(i)} \parallel \vec X^{(i)}$ to maximize $\ddot \tau$, and apply the maximal control amplitude $\lVert \vec h^{(i)} \rVert = h_\text{max}$, which is imposed by the experimental setup. Only when $\vec X^{(i)}$ vanishes within numerical accuracy, we will switch off the control Hamiltonian to regularize the problem.
 


 Typically, an initial state is prepared by a set of local measurements on the individual subsystems and is therefore separable. Although, in principle, arbitrary initial states can be chosen, in the specific case of the Ising Hamiltonian \eqref{eq:ising} that we will investigate below, eigenstates of $H_\text{sys}$ lead to a vanishing curvature. Consequently, we obtain vanishing partial derivatives \eqref{eq:partialderi} of $\ddot \tau$ with respect to the components of the control Hamiltonian $h^{(i)}_\xi$, \ie $\vec X ^{(i)} = 0$. In this case the method outlined above does not allow us to determine a suitable $H_\text{c}$, but an initial control pulse is readily applied to an eigenstate to obtain states of non-vanishing $\vec X ^{(i)}$.

 Since the partial derivatives $\partial \ddot \tau / \partial h_\xi^{(i)}$ in Eq.~\eqref{eq:partialderi} that make up $\vec X^{(i)}$ are algebraic expressions of the relevant system parameters (\ie $\vec X^{(i)} = \vec X^{(i)}(\rho,H_\text{sys},{\cal D})$), we have determined the optimal time-local control Hamiltonian by purely algebraic means. This offers a great advantage over numerical algorithms in optimal control theory, since for a given system we can express the control Hamiltonian $H_\text{c}$ in terms of an arbitrary state $\rho$, of the given intrinsic interaction $H_\text{sys}$, and of the given dissipator $\cal D$ once and for all. The optimal time evolution starting from an arbitrary initial state is then obtained by simple numerical integration, and, what is more, by one single propagation.

\begin{figure}[t]
   \centering
   \includegraphics[width=0.3\textwidth]{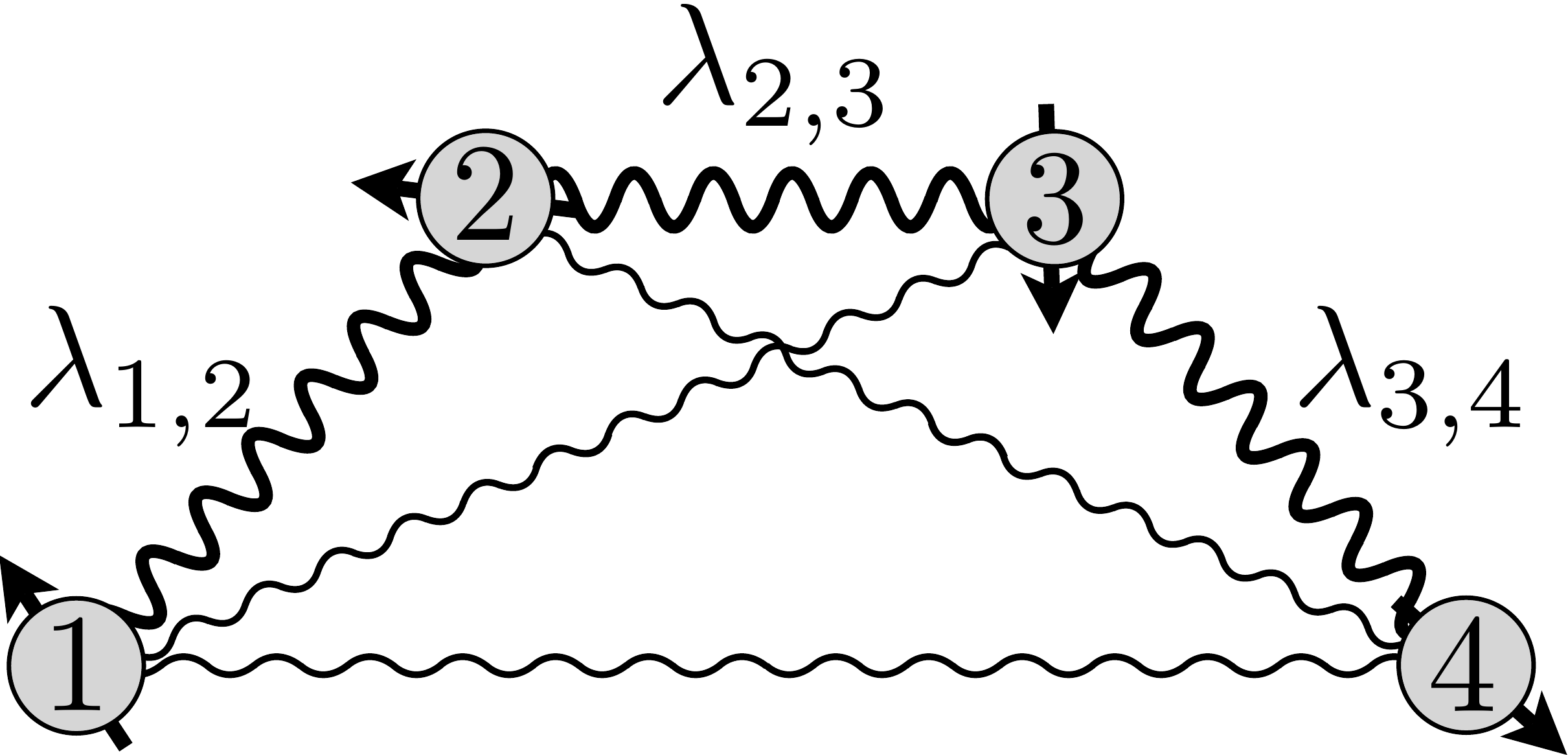}
   \caption{Four spins with permanent coupling---for example, nitrogen vacancy (NV) centers in diamond or nuclear spins in a molecule. Bold lines represent the dominant couplings as given by \eqref{eq:randomlambda}.}
   \label{fig:4NVcenters}
 \end{figure}

 \section{Application to four interacting particles: the coherent case}
 \label{sec:cohcontrol}
 
\begin{figure*}[t]
   \centering
   \includegraphics[width=0.4\textwidth]{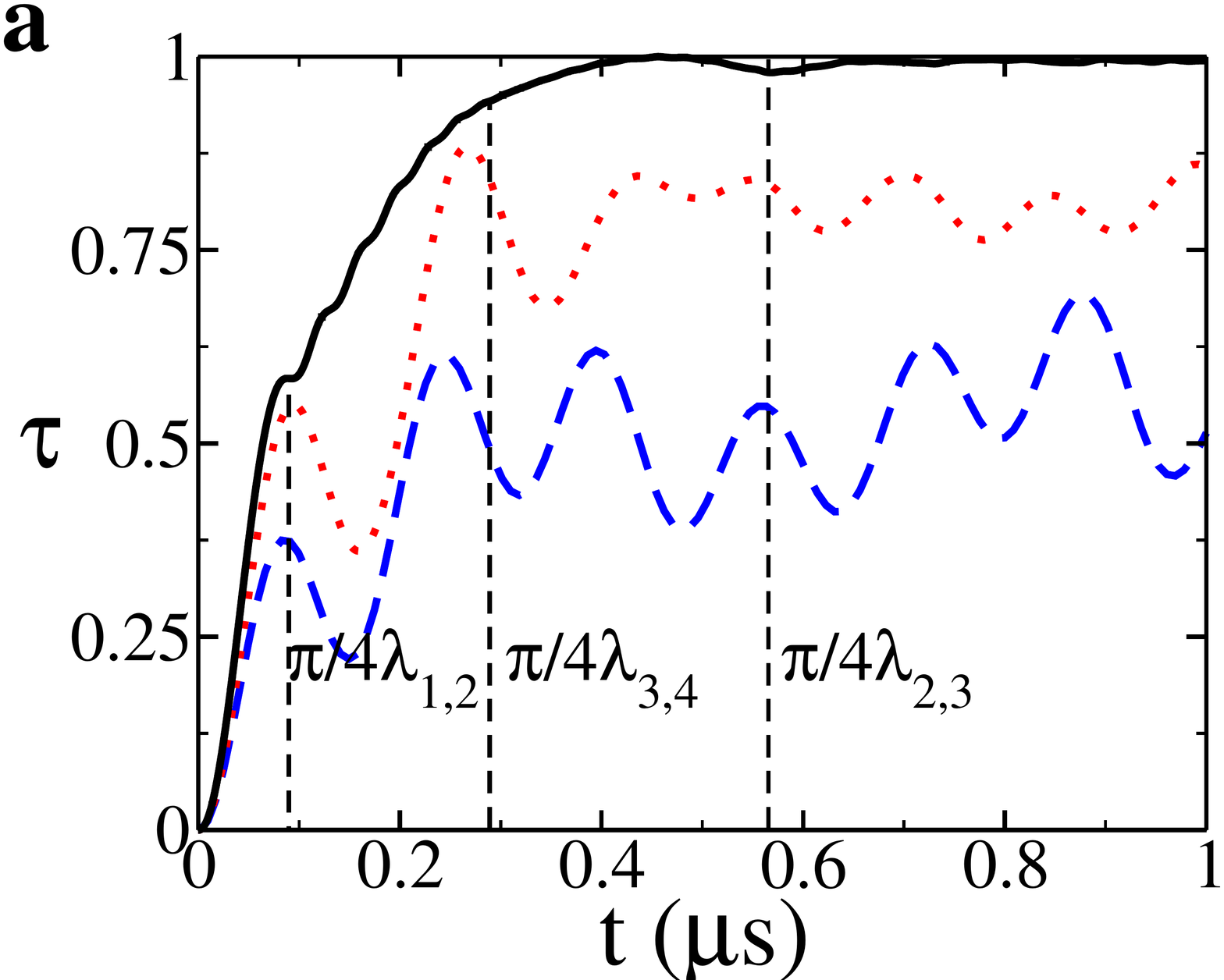}
   \includegraphics[width=0.4\textwidth]{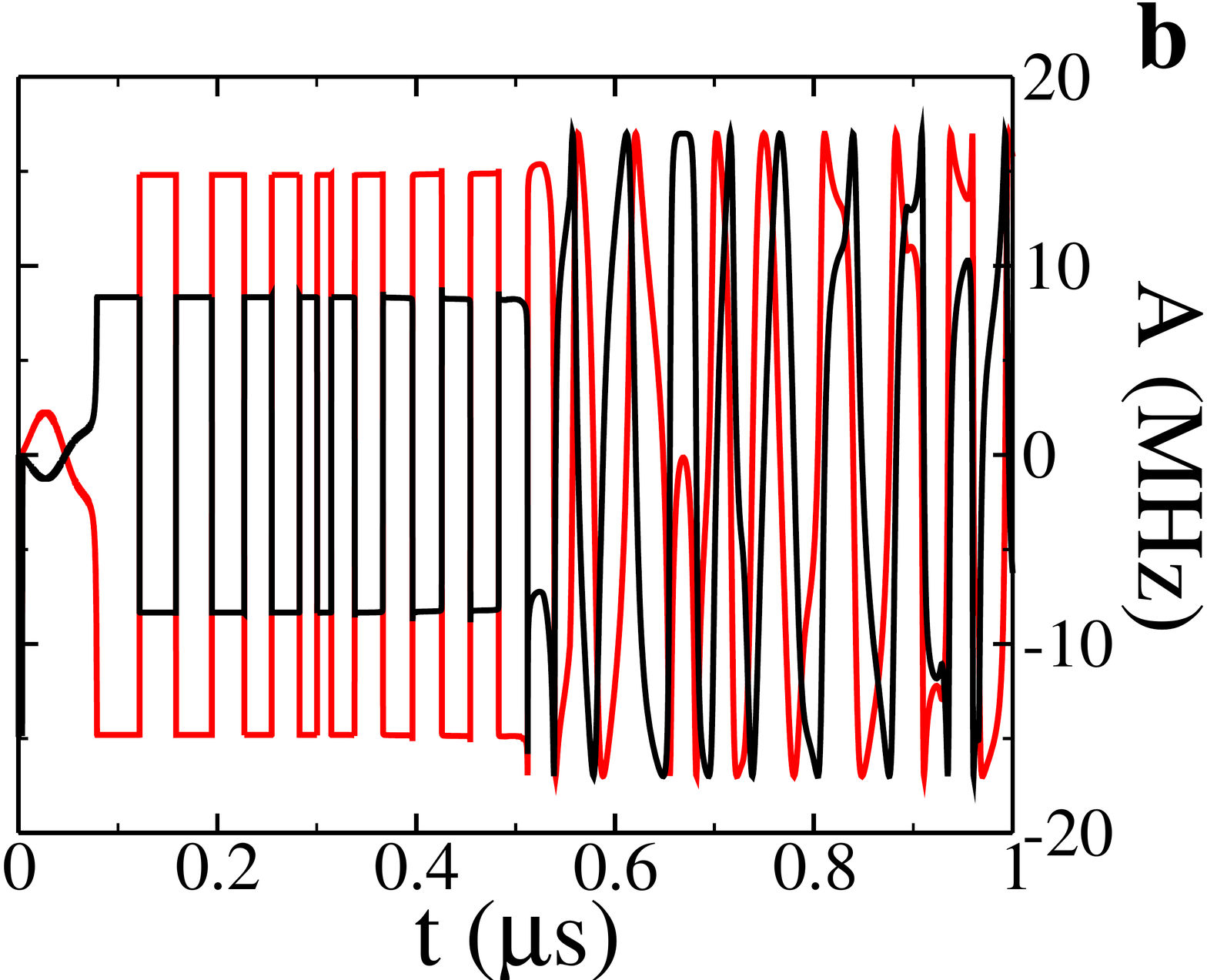}
   \caption{\textbf a---Time evolution of entanglement, quantified by the target functional $\tau$, for the separable initial state (\ref{eq:initialstate}) and coupling constants \eqref{eq:randomlambda}. The uncontrolled (dashed line) and optimized time evolution, with control amplitude 2.5~MHz (dotted line) and 17~MHz (solid line) are plotted. Characteristic time scales of $H_{\text{sys}}$ are indicated by dashed vertical lines. \textbf b---Optimal time-local control fields $h_x^{(1)}$ (black) and $h_y^{(1)}$ (red) addressing the first spin, with maximal control amplitude \mbox{$h_\text{max} = 17$ MHz}.}
   \label{fig:intensities}
\end{figure*}

After the derivation of the algebraic foundations of our control strategy, we now turn to the demonstration of its effectiveness and its robustness. For that purpose, let us consider four interacting spin-$1/2$ particles (Fig.~\ref{fig:4NVcenters}) with permanent interactions, as for example in nitrogen vacancy (NV) centers in diamond \cite{NVmanip} or nuclear spins in a molecule \cite{NMRqc}.

 In the present section we will investigate the \emph{coherent} time evolution (\ie in the absence of environment coupling)
under an Ising-type Hamiltonian
 \be
 	H_\text{sys} = \sum_{\substack{i,j=1 \\ i<j}}^4 \lambda_{i,j} \, \sigma_z^{(i)} \otimes \sigma_z^{(j)},
	\label{eq:ising}
 \ee
 as a model, for example, for dipole-dipole interactions. In the case of dipole-dipole interactions, the coupling constants $\lambda_{i,j}$ are fixed by the inter-particle distances,
and here the following randomly chosen set $\{\lambda_{i,j}\}$ is used:
 \begin{equation}
 \begin{matrix}
 	\lambda_{1,2} = \boldsymbol{9.8}\text{ MHz} & \lambda_{1,3} = 0.1\text{ MHz} & \lambda_{1,4} = 0.3\text{ MHz} \\
	& \lambda_{2,3} = \boldsymbol{1.3}\text{ MHz} & \lambda_{2,4} = 0.5\text{ MHz}\\
	&& \lambda_{3,4} = \boldsymbol{2.7}\text{ MHz}.\\
 \end{matrix}
 \label{eq:randomlambda}
 \end{equation}
 For the sake of concreteness, the interaction strengths are chosen to match the scale which is typically observed for NV centers \cite{Gaebel2006408}.
 
 Except for Sec.~\ref{sec:robuststates} we choose separable initial states to demonstrate the optimal creation of entanglement in the control system. The numerical results presented in the following have been obtained for the exemplary initial state
 \begin{equation}
 	\ket{\varphi_\text{i}} = \Big(\sqrt{\alpha}\ket 0 + \sqrt{1-\alpha}\;e^{i\theta}\;\ket 1 \Big)^{\otimes 4},
 	\label{eq:initialstate}
 \end{equation}
 with $\alpha= 0.73$ and $\theta = 0.51\pi$. Since local control rapidly drives the system towards that separable state with entanglement growing most rapidly the performance of our control scheme is largely independent of our choice of the separable initial state $\ket{\varphi_i}$, as we have also confirmed numerically.
 
 The time evolution of $\tau$ in the absence of control is depicted in Fig.~\ref{fig:intensities}~\textbf{a}, by a dashed line. $\tau$ starts out at zero, since the initial state is separable. Then, the entanglement increases due to interactions, and finally oscillates around mediocre values of about 0.5, on the characteristic time scales of $H_{\text{sys}}$ (indicated by dashed vertical lines). The most rapid oscillations of $\tau$ can be clearly identified with the time scale corresponding to $\lambda_{1,2}$, and they are superposed with oscillations on longer time scales corresponding to the other coupling constants. We can see that, on the considered time-interval of one micro-second, strong entanglement with $\tau \approx 1$ is not reached by the natural time evolution of the system.
  
 Let us now optimize the entanglement dynamics---\ie maximize the curvature of $\tau$---through continuous application of the optimal local control Hamiltonian derived in the previous section.
 The dotted line in Fig.~\ref{fig:intensities}~\textbf{a} shows the impact of weak control fields, with maximum amplitude that amounts to only a fraction of the system's intrinsic energy scale ($h_\text{max} = 0.26 \, \lambda_{1,2}$, where $\lambda_{1,2}$ is the largest coupling constant). We can see that, despite this rather small control amplitude, the system experiences a considerable increase of its average entanglement from 0.5 to about 0.8 after a time $t \approx \pi /4 \lambda_{3,4}$ (the second largest coupling constant). Yet, oscillations of $\tau$ prevail, since the natural dynamics is still dominant. 
 
 For the system dynamics becoming more and more dominated by the control Hamiltonian $H_\text{c}$, \ie for higher and higher control field amplitudes, Eq.~\eqref{eq:ddtauscalprod} predicts an increase of the curvature $\ddot \tau$ and therefore an increase of the average system entanglement. And indeed, our numerics confirm this stabilization of entanglement at ever higher average values when increasing the control amplitude. Still remarkably weak control fields with an amplitude of $17\text{ MHz} \approx 1.7\, \lambda_{1,2}$ (solid line in Fig.~\ref{fig:intensities}~\textbf{a}) result in maximal entanglement $\tau =1$, after a control time of the order of $t=\pi / 4 \lambda_{2,3}$, where $\lambda_{2,3}$ is the third largest coupling constant. Oscillations of $\tau$ are flattened out, and once maximal entanglement is reached it is essentially maintained.
 
\begin{figure}[t]
   \centering
   \includegraphics[width=0.4\textwidth]{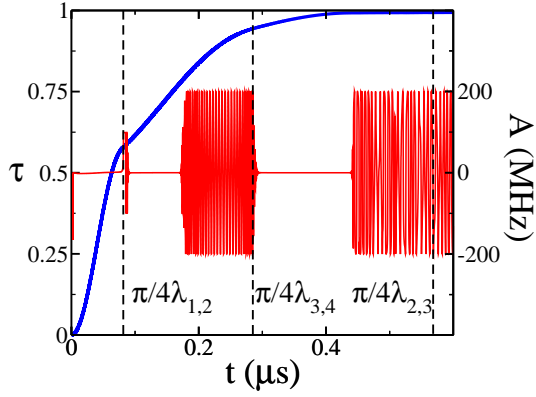}
   \caption{Time evolution of $\tau$ (blue line) for the separable initial state \eqref{eq:initialstate}, under control with maximal intensity $h_\text{max} = 200$ MHz. The red line represents the $h_x^{(1)}$-component of the applied control field (axis on the right).}
 \label{fig:int200}
 \end{figure}

 However, further augmenting the control amplitude yields no substantial acceleration of entanglement creation, and saturation sets in. The saturation of $\tau$ is apparent for example for $h_\text{max} = 200 \text{ MHz} \approx 20\, \lambda_{1,2}$, as depicted in Fig.~\ref{fig:int200} by a blue line. Now, $\tau(t)$ is perfectly smooth, but reaches its maximum essentially at the same time as for control with an amplitude of 17 MHz. This saturation is due to the fact that a local control Hamiltonian on its own {\em cannot} create any entanglement, but can only help to exploit the interactions in an optimal fashion. It is the interaction strength that sets the limiting timescale for the creation of entanglement.
 
 Let us discuss a bit more in detail how our applied control fields achieve the maximization of entanglement. First, note that the control fields act on the time scale associated with the Rabi period of their amplitudes $h_\text{max}$: \eg in the case of optimal control with $h_\text{max} = 17$ MHz (solid line in Fig.~\ref{fig:intensities}~\textbf{a}), the control fields $h_x^{(1)}$ and $h_y^{(1)}$ (see Fig.~\ref{fig:intensities}~\textbf{b}) oscillate with a frequency of  17 MHz. This is due to the fact that, on the one hand, $H_\text{c}$ induces a change in $\rho$ on this time scale and, on the other, that $H_\text{c}$ is implicitly time dependent through its dependence on $\rho(t)$.

 Therefore, as we increase the control field amplitude we also change the time scale on which our control fields can react to the natural system dynamics. For values of $h_\text{max}= 200$ MHz,
we expect that the time scales of $H_\text{c}$ and $H_\text{sys}$ separate, and that the control Hamiltonian can react instantly to the natural time evolution of the system. Indeed, instantaneous control is manifest in sharp pulses in the $h_x^{(1)}$-component of the control field at $t=0$ and $t=\pi/4\lambda_{1,2}$ (see Fig.~\ref{fig:int200}, red line, y-axis on the right hand side). They rapidly drive the system to those states whose entanglement subsequently increases during a finite time interval under $H_\text{sys}$ alone, with no external control present. In addition, during a time interval before $t = \pi/4\lambda_{3,4}$ and after reaching maximum entanglement, it seems to be necessary to supplement the natural dynamics with permanent control. In particular, once maximal entanglement is reached after $t\approx 0.45~\mu$s, $H_\text{c}$ and $H_\text{sys}$ act as an effective Hamiltonian that performs only local unitaries on the final state $\ket{\varphi_\text{f}}$, as will be verified shortly, thus maintaining the value of entanglement at $\tau = 1$.

 Second, let us turn to the efficiency of the control process. As argued above, the only mechanism that can create entanglement in the considered setup is the interaction. It therefore defines a limiting time scale $t_\text{E}$ below which maximal entanglement cannot be created. The greatest coupling constant in $H_\text{sys}$ ($\lambda_{1,2}$ in the case of coupling constants \eqref{eq:randomlambda}) defines the time-scale at which one Bell pair can be created. In the present case this Bell pair is formed between qubits one and two. And indeed, around the kink in the evolution of $\tau(t)$ at $t = \pi/4\lambda_{1,2}$ (as seen Fig.~\ref{fig:int200} and, less prominently, in Fig.~\ref{fig:intensities}) the entanglement corresponds to that of a single Bell pair, namely $\tau=1/2$. Analogous arguments apply for the second largest coupling constant (in this case $\lambda_{3,4}$), and indicate that, after $t =\pi/4\lambda_{3,4}$, the system can be described as a pair of Bell pairs. Finally, the third strongest interaction entangles the two Bell-pairs and creates genuine four-particle entanglement, and maximum entanglement is reached on the time scale of the third largest coupling constant $t_\text{E} = \pi / 4 \lambda_{2,3} $.

 In other words, each pairwise interaction leads to quantum correlations between two constituents and three pairwise couplings are needed to connect the four particles in our setup. Therefore, it is not surprising that genuine four-particle entanglement (\ie four particle quantum correlations) can only arise on the time scale of the smallest of these three coupling constants. We want to stress here, that, with our control strategy, maximum entanglement is reached on this shortest possible time scale.
 
 At the end of the control process, our analyses indicate that the target functional 
 $\tau$ in Eq.~\eqref{eq:mpconc2} is maximized by final states that are equivalent to
 \be
 	\ket{\varphi_\text{max}} = \frac 1 2 \Big( \ket{0000} + \ket{1111} + i(\ket{0011} + \ket{1100}) \Big)
	\label{eq:maxstate}
 \ee
 under local rotations or possibly permutations of subsystems. This has been verified for a large number of different initial states and coupling constants. Interestingly, the specific ordering of subsystems in the final state of the control process is determined by the largest coupling constants: if qubits $i$ and $j$ are strongly coupled, those states of maximal $\tau$ will be preferred that involve the collective ground and excited states of the pair $(i,j)$.
 In the present case of a system Hamiltonian with coupling constants given by \eqref{eq:randomlambda}, where qubits one and two and qubits three and four are particularly strongly coupled, we observe the creation of the collective ground and excited state of the pair $(1,2)$ and of the pair $(3,4)$---\ie the state given in Eq.~\eqref{eq:maxstate}---up to local rotations.
 
 \begin{figure}[t]
   \centering
   \includegraphics[width=0.4\textwidth]{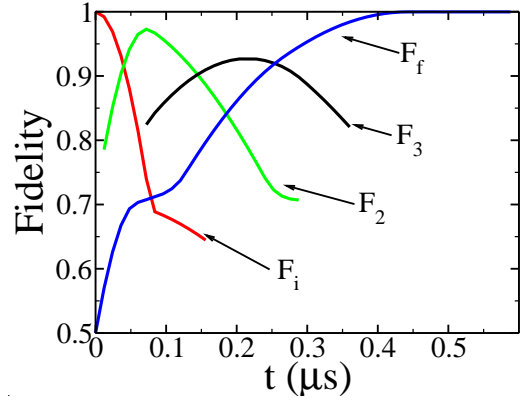}
   \caption{Dynamical entanglement build up underlying Fig.~\ref{fig:int200}: the system evolves from a separable initial state $\ket{\varphi_\text{i}}$ into a two-particle entangled state $\ket{\varphi_2}$, then into a pair of entangled Bell-pairs $\ket{\varphi_3}$, and finally into the genuine four-particle entangled state $\ket{\varphi_\text{f}}$. The plot shows the absolute value of the overlaps of the system's state $\ket{\psi(t)}$ with these four states, maximized over local unitaries---\ie $F_j(t)$ defined in Eq.~\eqref{eq:maximizedoverlaps}.}
 \label{fig:fidelities}
 \end{figure}
 
 To summarize, we expect $\ket{\psi(t)}$ to evolve into the genuinely four-particle entangled state $\ket{\varphi_\text{max}}$ in a stepwise process: from a separable initial state, via a two-particle entangled state, and a state of two entangled pairs, towards $\ket{\varphi_\text{max}}$
 \begin{multline}
   \ket{\varphi_\text{i}} = \frac 1 4 (\ket 0 + \ket 1)^{\otimes 4} \\
    \shoveleft{\begin{aligned}
	\stackrel{\lambda_{1,2}}{\longrightarrow} \ket{\varphi_2} &= \frac {1} {2\sqrt 2} (\ket {00} + \ket {11}) \otimes (\ket 0 + \ket 1)^{\otimes 2}\\
	\stackrel{\lambda_{3,4}}{\longrightarrow} \ket{\varphi_3} &= \frac 1 2 (\ket {00} + \ket {11})^{\otimes 2}\\
	\stackrel{\lambda_{2,3}}{\longrightarrow} \ket{\varphi_\text{f}} &= \ket{\varphi_\text{max}}
 \end{aligned} }\\
 =\frac 1 2 \Big(\ket{0000} + \ket{1111} + i(\ket{1100} + \ket{0011})\Big),
\label{eq:optimalpath} 
 \end{multline}
 where all states are given up to local unitary transformations.
 
 The sequential build-up of the many-body entangled state in \eqref{eq:optimalpath} is explicitly retraced in Fig.~\ref{fig:fidelities}, where we plot
 \begin{gather}
  F_j(t) =\max_{{\cal U}_1,{\cal U}_2,{\cal U}_3,{\cal U}_4} \left|\bra{\varphi_j}{\cal U}_1\otimes{\cal U}_2\otimes{\cal U}_3\otimes{\cal U}_4\ket{\Psi(t)}\right|\ ,\label{eq:maximizedoverlaps} \\
  \text{with } j \in \{\text{i},2,3,\text{f}\}, \nonumber
 \end{gather}
 \ie the maximal overlap of $\ket{\Psi(t)}$ with states that are equivalent to $\ket{\varphi_j}$ under local unitaries. It shows a consecutive emergence of maxima of the $F_i(t)$ in exactly the order that was predicted in \eqref{eq:optimalpath}. The fact that $|\ovl{\varphi_\text{f}}{\psi(t)}|$, with the above maximization, reaches a value of one and is then maintained at this value, implies that the joint action of $H_\text{sys}$ and $H_\text{c}$ reduces to that of a local unitary after the final state $\ket{\varphi_\text{f}}$ is reached.

 \section{Application to four interacting particles: the incoherent case}
\label{sec:dissipativesystems}

 Whereas in the previous section we have considered the idealized case of a closed quantum system, in order to obtain a basic understanding of entanglement dynamics under optimal time-local control, we have to bear in mind that in real world experiments it is never possible to shield a system completely from environmental effects. Therefore, it is mandatory to consider the control of \emph{open systems}. As mentioned before, our control Hamiltonian was designed to explicitly account for decoherence and, what is more, to minimize its detrimental impact on entanglement.
 
 Here, we will focus on the paradigmatic model of a dephasing environment. In particular, we will consider the situation where each spin is coupled to a separate environment, meaning that spins cannot interact with each other via the environment. This assumption is justified if the environmental noise at the location of different spins is completely uncorrelated, \eg due to their big spatial separation, or due to the sufficiently rapid decay of bath correlation functions. In this case, the Lindblad operators in the master equation act locally, and we have $L_i = \sigma_z^{(i)}$ in Eq.~\eqref{eq:drho} (\ie the third Pauli matrix, acting on the $i^\text{th}$ spin). Furthermore, we will assume that the dephasing rates felt by different spins are equal ($\gamma_i = \Gamma$, for $i = 1 \ldots 4$).
 
\begin{figure}[t]
   \centering
   \includegraphics[width=0.4\textwidth]{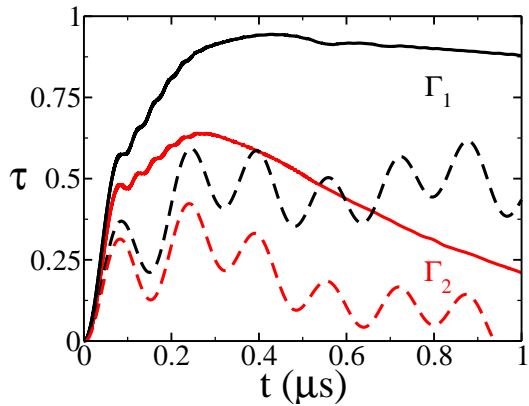}
   \caption{Time evolution of entanglement of four interacting spins described by (\ref{eq:randomlambda}, \ref{eq:initialstate}), under the influence of dephasing with rates $\Gamma_1=0.02 \, \mu \text{s}^{-1}$ (black lines) and $\Gamma_2=0.2 \, \mu \text{s}^{-1}$ (red lines): evolution of a separable initial state without control (dashed lines) and under control with maximal amplitude 17~MHz (solid lines).}
   \label{fig:diss}
\end{figure}

 We first consider four spins that experience a dephasing rate of $\Gamma_1 = 0.02 \, \mu \text{s}^{-1}$ and interact through dipole-dipole interactions quantified by \eqref{eq:randomlambda}. As the interaction strengths \eqref{eq:randomlambda} are roughly based on the time scales realized in NV centers, it has to be mentioned that $\Gamma_1$ is two orders of magnitude stronger than decoherence rates in state of the art NV center experiments \cite{NVcohtime}. Since $\Gamma_1^{-1} = 50 \mu$s, the average life-time of single-particle coherences in the density matrix $\rho$ is about two orders of magnitude longer than the characteristic time scale of the system Hamiltonian, which is of the order of $0.1 \mu$s and we expect the time-evolution of $\rho$ to be dominated by interactions. One has to keep in mind though, that genuine multipartite entanglement relies on the existence of \emph{many-body} coherences which generally fall off much more rapidly as the number of particles involved increases \footnote{This can be seen by analytically solving Eq.~\eqref{eq:drho} for $L_i = \sigma_z^{(i)}$ and $H_\text{c} = H_\text{sys}=0$.}.
 
 Fig.~\ref{fig:diss} shows the time evolution of $\tau$ under the influence of dephasing with rate $\Gamma_1$. The free entanglement dynamics in the absence of control (dashed line) is reminiscent of the purely coherent case (Fig.~\ref{fig:intensities}, dashed line), with oscillations on time scales of $H_\text{sys}$ around $\tau(t) \approx 0.5$. As expected, coherent oscillations between states that bear different amounts of entanglement prevail over the entire observation period of one microsecond.
 
 In contrast, our control scheme steers the system into almost maximal entanglement, $\tau = 0.95$ (black solid line), for the maximum control amplitude $h_\text{max} = 1.7 \lambda_{1,2}$.
 And this is achieved despite the fact that our entanglement measure targets genuine four-particle entanglement, which, as mentioned above, is most prone to environmental noise. Although decoherence cannot be avoided completely and entanglement will subsequently decay, the controlled entanglement dynamics outperforms the uncontrolled time evolution by far.
 
 In order to test fundamental limits for the creation and existence of entanglement we imposed even stronger dephasing with rate $\Gamma_2 = 0.2 \, \mu \text{s}^{-1}$ (red lines in Fig.~\ref{fig:diss}). Even in this case, our control strategy creates a considerable amount of entanglement (solid line). Despite the fact that dephasing takes its toll on entanglement as time goes by, the control fields stabilize the amount of entanglement at an average value which almost doubles that achieved by the uncontrolled system dynamics (dashed line).

\section{Robust entangled states}
\label{sec:robuststates}

 We have seen that our control strategy permits to create considerable entanglement in spite of the detrimental influence of dephasing. This efficiency in the presence of environment coupling is a consequence of the maximization of $\ddot\tau$ in Eq.~\eqref{eq:ddtauscalprod}. Since decoherence effects explicitly enter this expression, we claimed that the system is steered into the state that is least sensitive to the dissipator $\mathcal D (\rho)$, \ie most robust against environment coupling. This claim will be substantiated in the following.

 As argued above, the open system evolution considered up to now was dominated by the system's coherent dipole-dipole interactions, and decoherence only made itself felt through a gradual degradation of entanglement. A state's robustness with respect to decoherence, however, becomes most obvious if we consider systems with negligible intrinsic interactions (\eg if the spins are located a large distances), such that entanglement is expected to decay monotonically \cite{PhysRevLett.92.180403,flodeco}. 

\begin{figure}[!t]
   \centering
   \includegraphics[width=0.4\textwidth]{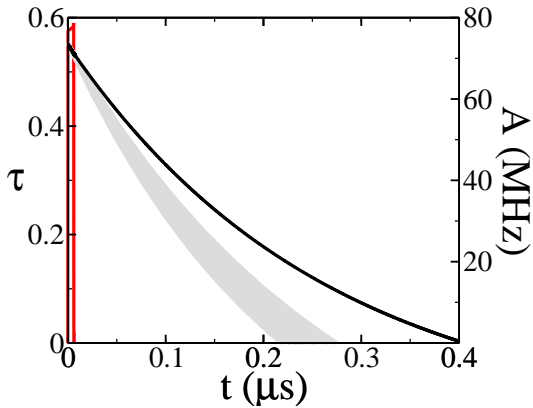}
   \caption{Entanglement evolution in a non-interacting system under dephasing with rate $\Gamma = 0.8 \, \mu \text{s}^{-1}$. We compare $\tau(t)$ for a pure, entangled initial state under control (black line, $h_\text{max} = 17$ MHz) with that for 50 local unitary equivalent states without control (grey area). The optimal time-local control Hamiltonian (here the exemplary $h_x^{(1)}$-component) is comprised of only one short pulse in the beginning (red line, axis on the right).
   }
   \label{fig:robustdeco}
 \end{figure}

 Consequently, we now assume that the spins be uncoupled ($\lambda_{i,j} = 0, \forall \, i,j$), but subject to dephasing with rate $\Gamma = 0.8 \, \mu \text{s}^{-1}$. Additionally, we take a pure, entangled initial state with $\tau(0) = 0.55$, and compare its entanglement evolution under control (black line in Fig.~\ref{fig:robustdeco}) with the entanglement evolution of 50 uncontrolled, local-unitary equivalent initial states, that are uniformly distributed with respect to the Haar measure \cite{haarmeasure} (gray area). And indeed, the control fields produce that state which is most robust to decoherence, with an entanglement lifetime about 1.75 times longer than that of generic, uncontrolled and therefore non-robust states. This advantage of the most robust state over non-robust states in terms of entanglement lifetime is less pronounced for higher initial entanglement: for an initial state with $\tau(0) = 0.8$ ($\tau(0) = 1$), the time span after which entanglement has decayed is prolonged only by a factor of 1.3 (1.05). This, however, is not due to an insufficiency of our control strategy---it finds the most robust states, irrespective of the initial entanglement.

 The existence of particularly robust entanglement among weakly entangled states has implications both for fundamental and practical considerations. On the one hand, it may shed light on the possibility for entanglement to survive even in situations of extreme environment coupling \cite{ISI:000279014400020,2011arXiv1104.3883T}---a situation that strongly contrasts our intuition, which is based on experiments with highly entangled states~\cite{PhysRevLett.106.130506}. On the other hand, when it comes to the exploitation of entanglement for quantum information processing, weakly entangled states can be more useful than highly entangled ones \cite{gross:190501}, which underlines the potential that lies in the choice of states with particularly robust entanglement properties.
 


 
 Beyond the existence of particularly robust, weakly entangled states, there are two more facts that strike the eye in the above control setting with dephasing. First, the control Hamiltonian (its $h^{(1)}_x$-amplitude is represented by a red line in Fig.~\ref{fig:robustdeco} with the corresponding $y$-axis on the right) mainly consists of a short pulse at the beginning of the time evolution, and only very weak control thereafter (not discernible on the scale of the plot). This implies that the main contribution to the robustness of entanglement lies in choosing a proper initial state, and that its robustness is conserved under the dynamics of dephasing. Second, the robust state found by our control strategy is \emph{outstandingly} robust: its entanglement sticks out significantly from that of a generic uncontrolled state and among the 50 shown in Fig.~\ref{fig:robustdeco} none comes even close. This indicates that robustness is a property that is shared by only very few states. We have, however, ruled out that they merely form a set of measure zero within state space, by verifying that their robustness is insensitive to perturbations in first order. More specifically, a local unitary rotation of the robust state by 0.01$\pi$ (0.05$\pi$) leads to at most 0.1\% (1.6\%) higher entanglement decay rates. This means that robustness will not be destroyed by small, unavoidable experimental imperfections.
 
 \section{Robustness of control pulses}
 \label{sec:pulserobustness}
 
 \begin{figure*}[t]
   \centering
  \includegraphics[width=0.4\textwidth]{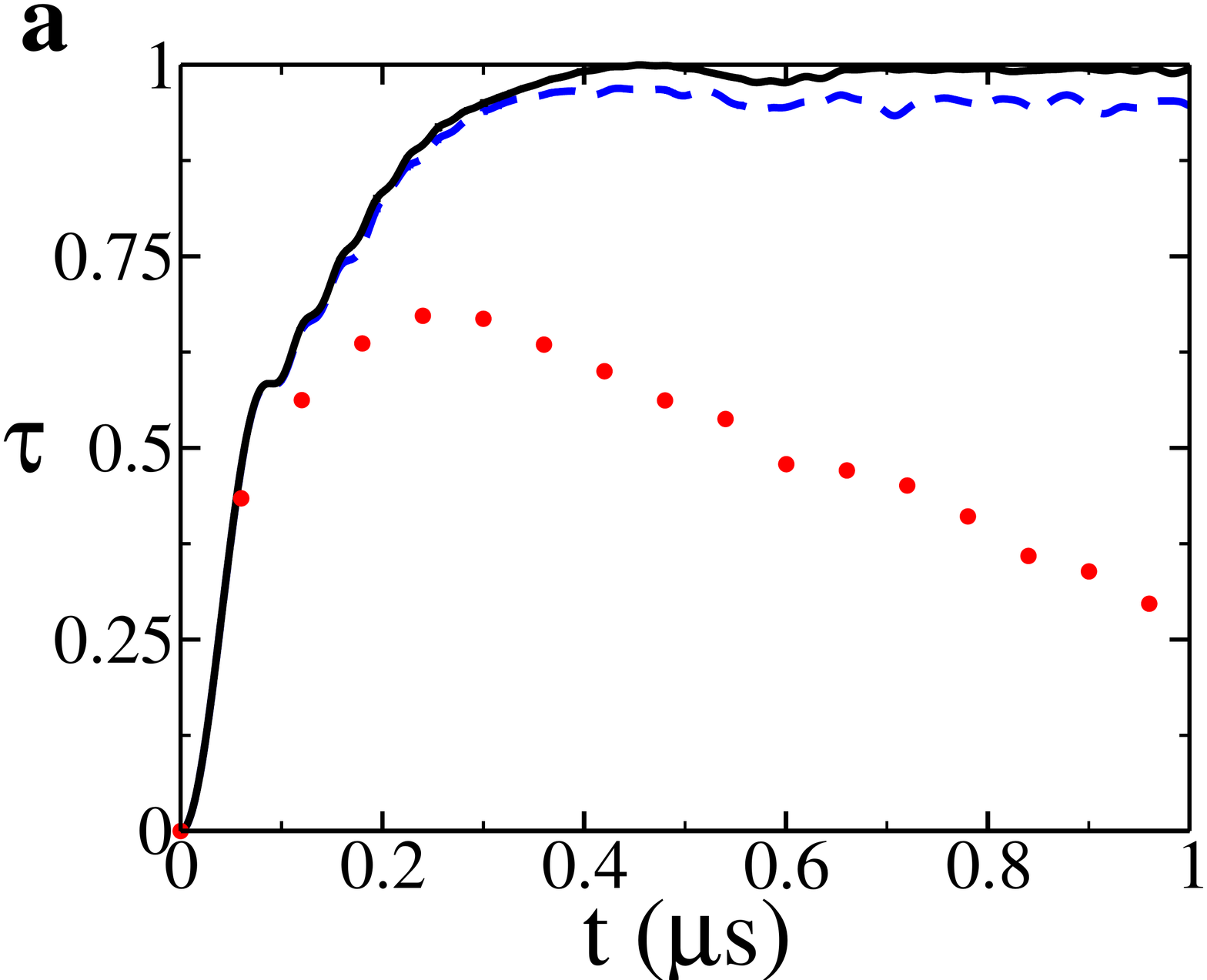}
  \includegraphics[width=0.4\textwidth]{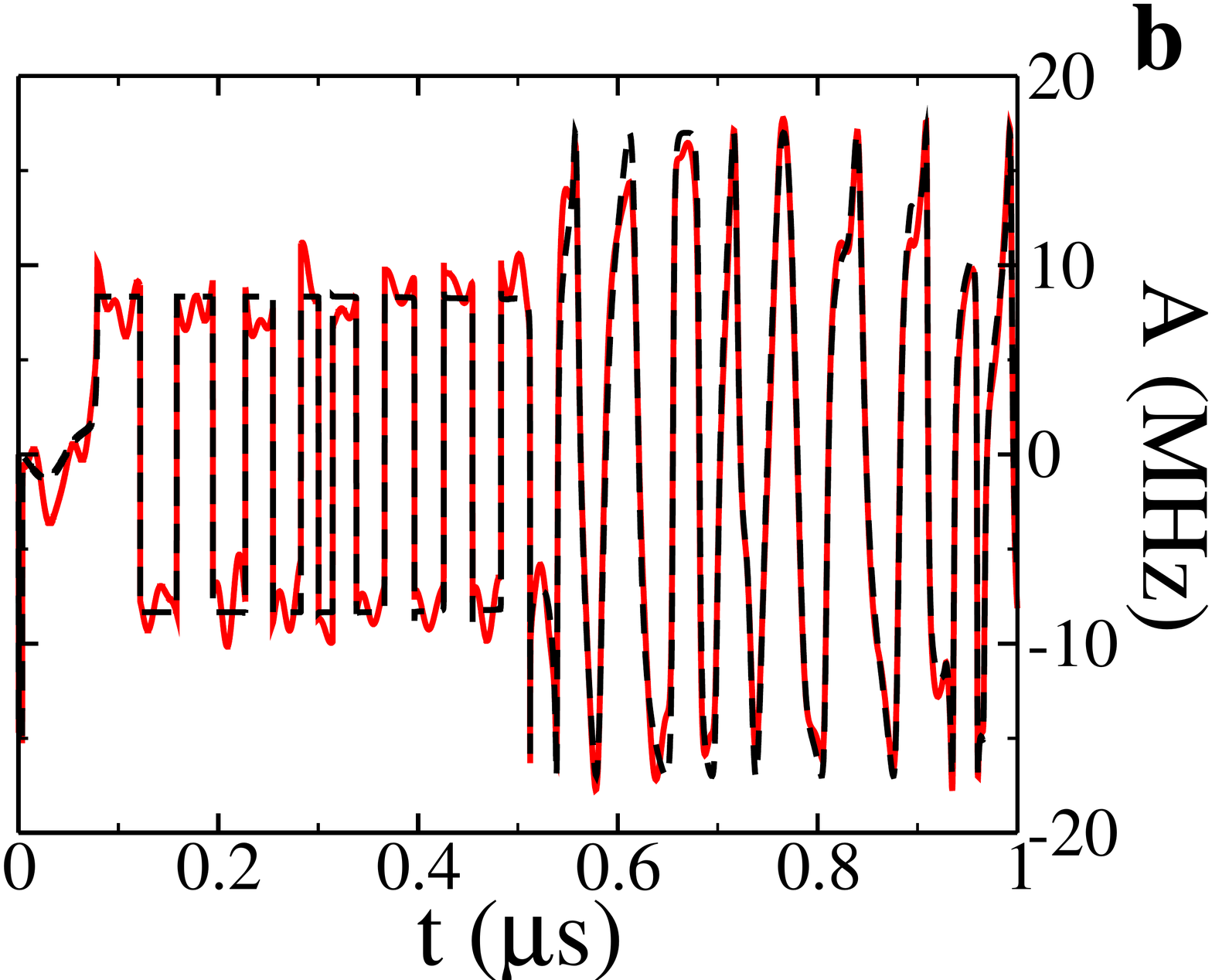}
   \caption{\textbf{a}---$\tau(t)$ under unperturbed and perturbed control with 10\% white noise in the control-pulses (black and blue dashed lines, respectively). Red dots represent the entanglement of the probabilistic mixture of 200 runs, with different realizations of white noise. \textbf{b}---Unperturbed and perturbed control field $h_x^{(1)}$ (black and red lines, respectively) corresponding to \textbf{a}.}
   \label{fig:whitenoise}
 \end{figure*}
 
 Real world experiments on quantum systems not only suffer from environment coupling, as considered above, but also from experimental imperfections, such as amplitude or phase fluctuations of the control laser fields, or imprecise knowledge of system properties, such as the coupling constants $\lambda_{ij}$ in \eqref{eq:randomlambda}. A very important aspect of the performance of control sequences in general is therefore their robustness with respect to such imperfections. In this section we will test the performance of the optimal control pulses under white noise perturbations of the control fields, constant offsets of their amplitudes, and imprecisely estimated coupling constants.
 
 To estimate the impact of experimental imperfections on entanglement correctly, it is crucial to distinguish between systematic and random (or stochastic) errors. Whereas the former can be assumed constant in consecutive runs of the experiment, the latter will typically differ from one run to another. Taking into account that the quantification of entanglement, as that of any physical property, requires a whole series of measurements on consecutively prepared system states,
 the final state in a randomly perturbed control process needs to be described as a probabilistic mixture of pure states $\{\ket{\psi_i}\}$, \ie a mixed state $\rho$. According to Eq.~\eqref{eq:convexroof}, the entanglement inscribed in the mixed final state $\rho$ is then generally lower than the average entanglement in the pure state ensemble $\{\ket{\psi_i}\}$. In the presence of a systematic error, however, one will prepare one and the same faulty final pure state in consecutive runs. Hence, the final entanglement will not suffer from an additional final state uncertainty, and can be estimated as the average pure state entanglement for different realizations of the systematic error.

 In order to quantify the influence of perturbations or parameter uncertainties with relative error $\varepsilon$, we compare the perturbed and unperturbed evolution of the system already treated in section~\ref{sec:cohcontrol}---\ie four spins with dipole-dipole interactions as specified by Eqs.~(\ref{eq:randomlambda}, \ref{eq:initialstate}), in the absence of dephasing. First, let us consider the random white noise perturbation of the control fields. In Fig.~\ref{fig:whitenoise}~\textbf{a} the entanglement dynamics is plotted for a single run of the control process, perturbed by a white noise signal (flat spectrum, cutoff at $100 \pi$ MHz) with 10\% relative amplitude, \ie ${\varepsilon = 0.1}$ (blue dashed line). Compared to the unperturbed time evolution of $\tau$, the maximal entanglement is reduced by only 5\%, and this deviation slowly increases with time. The unperturbed and the perturbed control field for this specific realization can be compared in Fig.~\ref{fig:whitenoise}~\textbf{b}. The entanglement of the probabilistic mixture of the states prepared in many different runs with different realizations of white noise at equal relative amplitude converges to the red dotted line in Fig.~\ref{fig:whitenoise}~\textbf{a}. We can see that, as expected, it falls off significantly below the pure state entanglement in a single run.

\begin{figure*}[t]
   \centering
   \includegraphics[width=0.387\textwidth]{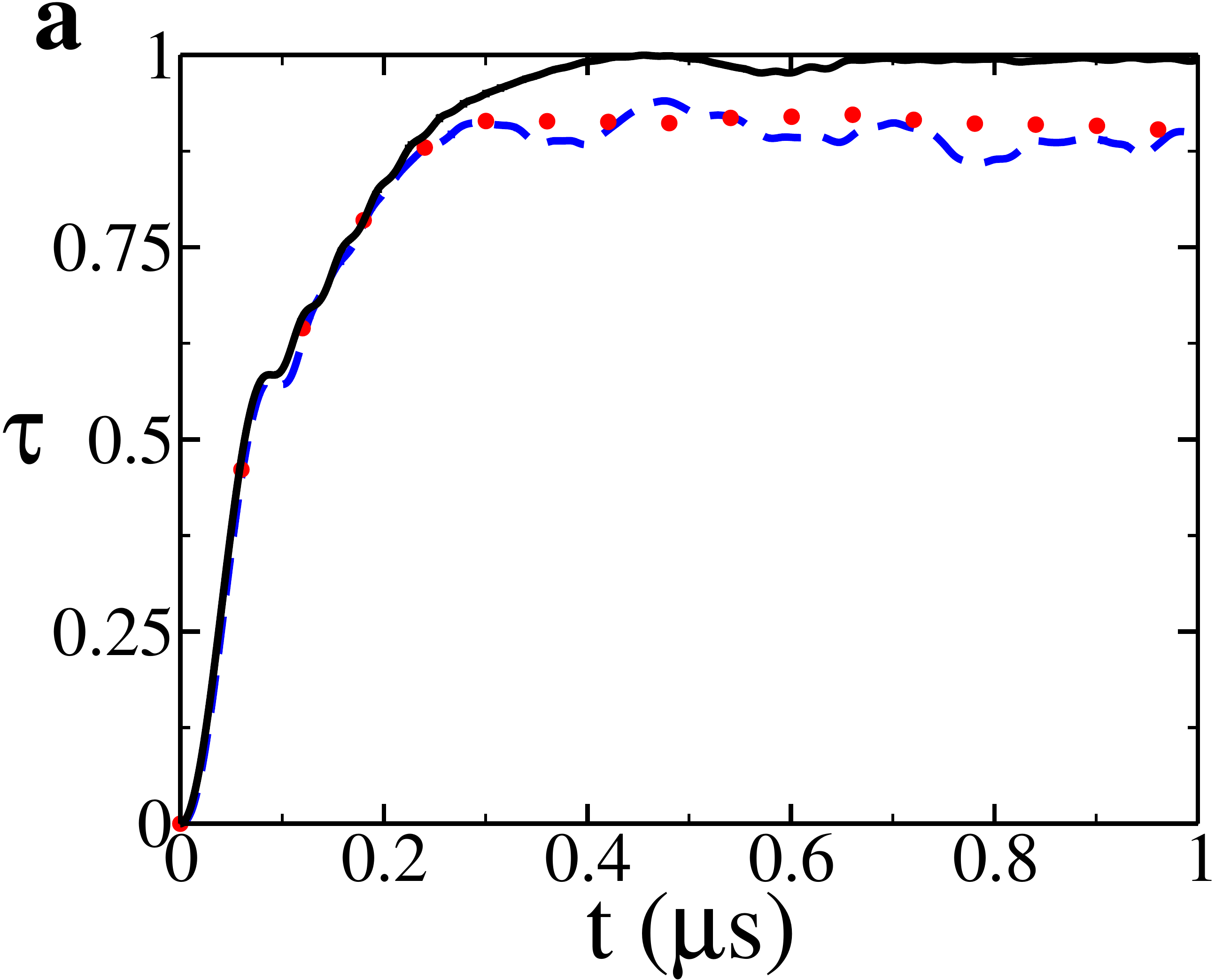} \qquad
   \includegraphics[width=0.379\textwidth]{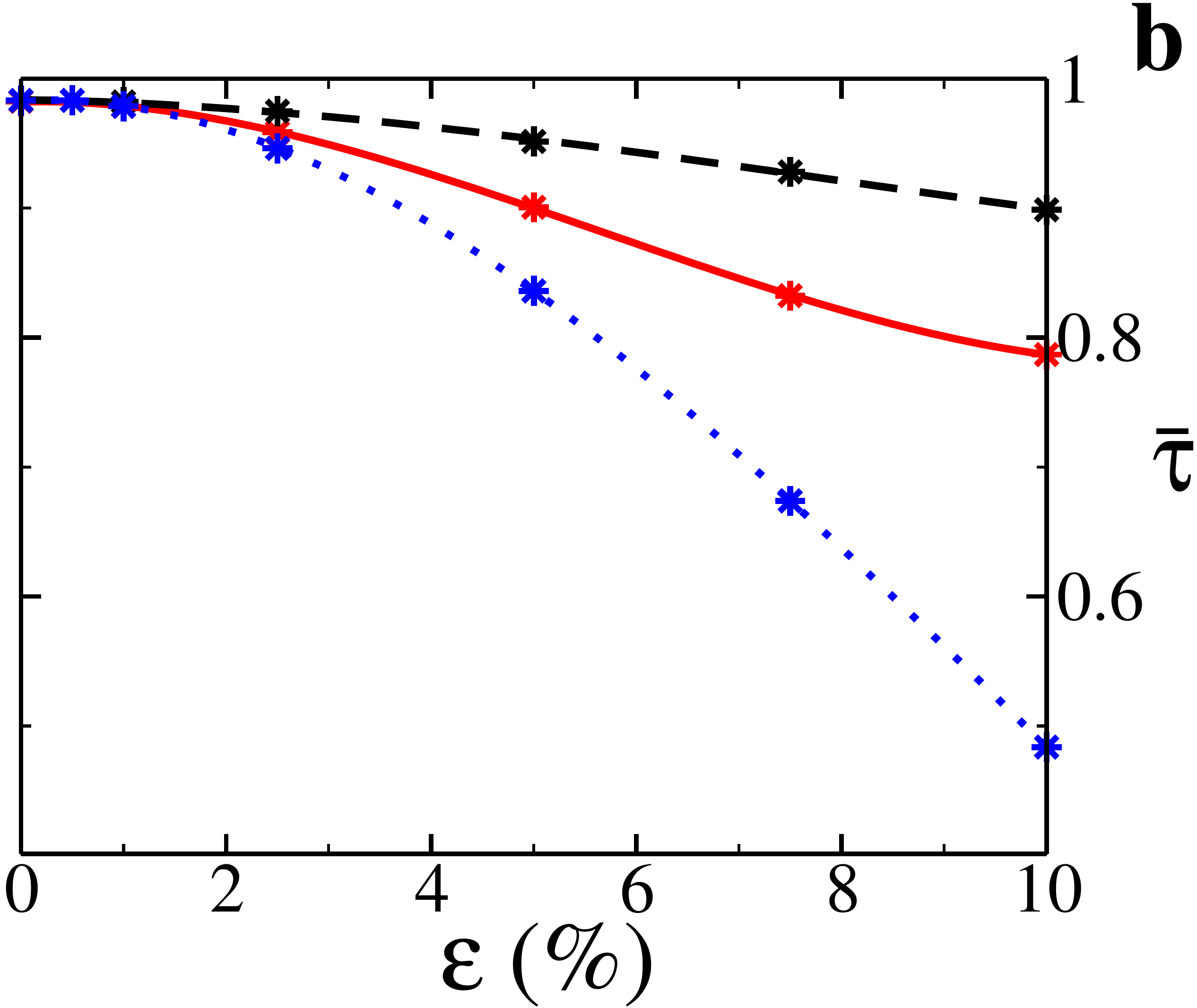}
   \caption{
 \textbf{a}---$\tau(t)$ under control and in the presence of a 10\% uncertainty in the dipole-dipole coupling constants $\lambda_{i,j}$ (blue dashed line), and averaged over 200 realizations of this 10\% error (red dotted line). \textbf{b}---Average maximum entanglement $\overline \tau$ reached by faulty control pulses with amplitude $h_\text{max}=17$MHz, averaged over 50 runs: $\overline \tau$ as a function of the relative error $\varepsilon$, for a white noise perturbation (blue dotted line), for a constant offset in the control fields (red solid line), and for uncertain coupling constants $\lambda$ (black dashed line).}
   \label{fig:constoff}
 \end{figure*}

 As an exemplary systematic error, we consider the case that the intrinsic dipole-dipole coupling constants $\lambda_{i,j}$ are not known with perfect accuracy, as it typically is the case due to measurement errors \cite{Gaebel2006408}. The entanglement dynamics in presence of a 10\% relative error of all $\lambda_{i,j}$, \ie $\varepsilon = 0.1$, is plotted in Fig.~\ref{fig:constoff}~\textbf{a}. Maximum entanglement is reduced by about 10\% in one single run, which is about twice as much as in the case of a single realization with white noise. As explained above, however, this systematic error will not induce mixing in the final state, and, hence, no additional penalty (in terms of entanglement) has to be paid upon experimental iteration. To rule out that, based on our specific realization of the error, we either over- or underestimate the impact on entanglement, we also averaged the entanglement dynamics over many runs with different realizations of the errors in $\lambda_{i,j}$ (red dotted line). The average entanglement dynamics confirms the 10\% reduction of entanglement.
 
 Finally, we want to quantify the entanglement drop-off as a function of the error parameter $\varepsilon$. Since the control Hamiltonian is chosen such that $\vec h$ and $\vec X$ are parallel to each other, an imperfection will typically result in a finite angle $\alpha$ between these two vectors. The curvature $\ddot\tau$ determined by the scalar product $\vec h \cdot \vec X$ (see Eq.~\eqref{eq:ddrho}) is consequently reduced by a factor of $\cos\alpha$, which will result in a slower increase of entanglement. Since, however, the $\cos$-function is independent of $\alpha$ in first order around $\alpha = 0$, we expect the performance of our control strategy to be insensitive to small imperfections.

 The performance for finite imperfections is shown in Fig.~\ref{fig:constoff}~\textbf{b}, where the average maximum entanglement $\bar \tau$ reached with perturbed control is depicted as a function of the relative strength $\varepsilon$ of the different error sources mentioned above (blue: white noise; black: wrong coupling constants $\lambda_{i,j}$, red: systematic offset of control fields). Conforming with our predictions on systematic and random errors, the white noise perturbation reduces the performance of the control process the most, whereas a systematic offset in the control fields and a wrong prediction of the coupling constants $\lambda_{i,j}$ have less negative impact.
 As expected, the first order term or slope at $\varepsilon = 0$ vanishes, \ie ${\left. \di \bar \tau / \di \varepsilon \right|_{\varepsilon = 0} = 0}$, for all three error-types. Even in the presence of errors in the percent-range, the system ends up in a strongly entangled state.
 The small offset $\bar \tau < 1$ for $\varepsilon =0$, where in the absence of errors one would have expected $\tau = 1$, arises since, in addition to an ensemble average, we also performed a time-average over the interval $0.3 \mu \text s \le t\le 1 \mu \text s$.
 This time-average, at the employed control amplitude of $h_\text{max} = 17$ MHz, comprises instances in which $\tau$ deviates from its maximum value, as can be seen for example in the black line in Fig.~\ref{fig:whitenoise}~\textbf{a} at $t\approx 0.6 \mu$s.

 In consequence, with the above robustness to systematic and random errors in first order and favorable scaling up to a few percents in the relative error, our control strategy proves to be suitable for experimental implementation.
 
\section{Conclusion}

We have seen that by applying a target functional with suitable invariance properties, quantum control is able to straightforwardly maximize dynamical quantities of many body systems. In the specific setting discussed here, we saw an essentially optimal exploitation of interactions for entanglement creation, and simultaneous mitigation of decoherence. In particular, this was achieved by time-local control, \ie without resorting to numerically expensive optimal control algorithms. However, the implementation of our approach is not linked to any particular control algorithm.

Moreover, with the target functional unburdened of unnecessary restrictions (\eg to specific system states), the control problem holds the potential to unveil dynamical system properties without the necessity for prior knowledge thereof. Here, we were able to identify the most robust states under environmental decoherence, among a multitude of equally entangled states. Needless to say that an investigation into their joint structural properties and relation to the predominant decoherence mechanism might be of great benefit for the creation of entanglement in open quantum systems.

 Our approach is by no means limited to the treatment of entanglement, as shown, for example,
by the successful use of invariant target functionals for quantum gates \cite{KochEntGate}, \ie elementary building blocks of quantum algorithms, or for the attenuation of polarization in nuclear magnetic resonance \cite{floglaser}. More generally, in complex quantum systems for which we lack a physical intuition, the use of control techniques that can identify states with particular properties bears the potential to fill this void. For example, the identification of states with particular robustness properties may substantially change our understanding of the existence of quantum signatures in hot, biomolecular processes such as photosynthesis \cite{engel,Scholes2011}. In quantum chemistry, one can also think of control tasks which target molecular (\ie many-body) properties associated not to specific electronic or vibrational states, but to a multitude thereof---think of bond length as a prominent example \cite{zewailbondbreak}. Instead of driving one specific pathway for a wave-packet on a landscape of Born-Oppenheimer surfaces, this could enable us to search for the most suitable quantum dynamics for the desired purpose, which, in general, involves complex superpositions in the high-dimensional Hilbert space of the molecule. In short, the present approach permits to extend the potential of quantum control by shifting the focus from reaching a goal in an optimal fashion to identifying the optimal goal in the first place.

\bibliography{biblio}

\begin{thebibliography}{40}%
\makeatletter
\providecommand \@ifxundefined [1]{%
 \@ifx{#1\undefined}
}%
\providecommand \@ifnum [1]{%
 \ifnum #1\expandafter \@firstoftwo
 \else \expandafter \@secondoftwo
 \fi
}%
\providecommand \@ifx [1]{%
 \ifx #1\expandafter \@firstoftwo
 \else \expandafter \@secondoftwo
 \fi
}%
\providecommand \natexlab [1]{#1}%
\providecommand \enquote  [1]{``#1''}%
\providecommand \bibnamefont  [1]{#1}%
\providecommand \bibfnamefont [1]{#1}%
\providecommand \citenamefont [1]{#1}%
\providecommand \href@noop [0]{\@secondoftwo}%
\providecommand \href [0]{\begingroup \@sanitize@url \@href}%
\providecommand \@href[1]{\@@startlink{#1}\@@href}%
\providecommand \@@href[1]{\endgroup#1\@@endlink}%
\providecommand \@sanitize@url [0]{\catcode `\\12\catcode `\$12\catcode
  `\&12\catcode `\#12\catcode `\^12\catcode `\_12\catcode `\%12\relax}%
\providecommand \@@startlink[1]{}%
\providecommand \@@endlink[0]{}%
\providecommand \url  [0]{\begingroup\@sanitize@url \@url }%
\providecommand \@url [1]{\endgroup\@href {#1}{\urlprefix }}%
\providecommand \urlprefix  [0]{URL }%
\providecommand \Eprint [0]{\href }%
\@ifxundefined \urlstyle {%
  \providecommand \doi  [0]{\begingroup \@sanitize@url \@doi}%
  \providecommand \@doi [1]{\endgroup \@@startlink {\doibase
  #1}doi:\discretionary {}{}{}#1\@@endlink }%
}{%
  \providecommand \doi  [0]{doi:\discretionary{}{}{}\begingroup
  \urlstyle{rm}\Url }%
}%
\providecommand \doibase [0]{http://dx.doi.org/}%
\providecommand \Doi [0]{\begingroup \@sanitize@url \@Doi }%
\providecommand \@Doi  [1]{\endgroup\@@startlink{\doibase#1}\@@Doi}%
\providecommand \@@Doi [1]{#1\@@endlink}%
\providecommand \selectlanguage [0]{\@gobble}%
\providecommand \bibinfo  [0]{\@secondoftwo}%
\providecommand \bibfield  [0]{\@secondoftwo}%
\providecommand \translation [1]{[#1]}%
\providecommand \BibitemOpen [0]{}%
\providecommand \bibitemStop [0]{}%
\providecommand \bibitemNoStop [0]{.\EOS\space}%
\providecommand \EOS [0]{\spacefactor3000\relax}%
\providecommand \BibitemShut  [1]{\csname bibitem#1\endcsname}%
\bibitem [{\citenamefont {Monz}\ \emph {et~al.}(2011)\citenamefont {Monz},
  \citenamefont {Schindler}, \citenamefont {Barreiro}, \citenamefont {Chwalla},
  \citenamefont {Nigg}, \citenamefont {Coish}, \citenamefont {Harlander},
  \citenamefont {H\"ansel}, \citenamefont {Hennrich},\ and\ \citenamefont
  {Blatt}}]{PhysRevLett.106.130506}%
  \BibitemOpen
  \bibfield  {author} {\bibinfo {author} {\bibfnamefont {T.}~\bibnamefont
  {Monz}}, \bibinfo {author} {\bibfnamefont {P.}~\bibnamefont {Schindler}},
  \bibinfo {author} {\bibfnamefont {J.~T.}\ \bibnamefont {Barreiro}}, \bibinfo
  {author} {\bibfnamefont {M.}~\bibnamefont {Chwalla}}, \bibinfo {author}
  {\bibfnamefont {D.}~\bibnamefont {Nigg}}, \bibinfo {author} {\bibfnamefont
  {W.~A.}\ \bibnamefont {Coish}}, \bibinfo {author} {\bibfnamefont
  {M.}~\bibnamefont {Harlander}}, \bibinfo {author} {\bibfnamefont
  {W.}~\bibnamefont {H\"ansel}}, \bibinfo {author} {\bibfnamefont
  {M.}~\bibnamefont {Hennrich}}, \ and\ \bibinfo {author} {\bibfnamefont
  {R.}~\bibnamefont {Blatt}},\ }\Doi {10.1103/PhysRevLett.106.130506}
  {\bibfield  {journal} {\bibinfo  {journal} {Phys. Rev. Lett.},\ }\textbf
  {\bibinfo {volume} {106}},\ \bibinfo {pages} {130506} (\bibinfo {year}
  {2011})}\BibitemShut {NoStop}%
\bibitem [{\citenamefont {Neumann}\ \emph {et~al.}(2008)\citenamefont
  {Neumann}, \citenamefont {Mizuochi}, \citenamefont {Rempp}, \citenamefont
  {Hemmer}, \citenamefont {Watanabe}, \citenamefont {Yamasaki}, \citenamefont
  {Jacques}, \citenamefont {Gaebel}, \citenamefont {Jelezko},\ and\
  \citenamefont {Wrachtrup}}]{Neumann06062008}%
  \BibitemOpen
  \bibfield  {author} {\bibinfo {author} {\bibfnamefont {P.}~\bibnamefont
  {Neumann}}, \bibinfo {author} {\bibfnamefont {N.}~\bibnamefont {Mizuochi}},
  \bibinfo {author} {\bibfnamefont {F.}~\bibnamefont {Rempp}}, \bibinfo
  {author} {\bibfnamefont {P.}~\bibnamefont {Hemmer}}, \bibinfo {author}
  {\bibfnamefont {H.}~\bibnamefont {Watanabe}}, \bibinfo {author}
  {\bibfnamefont {S.}~\bibnamefont {Yamasaki}}, \bibinfo {author}
  {\bibfnamefont {V.}~\bibnamefont {Jacques}}, \bibinfo {author} {\bibfnamefont
  {T.}~\bibnamefont {Gaebel}}, \bibinfo {author} {\bibfnamefont
  {F.}~\bibnamefont {Jelezko}}, \ and\ \bibinfo {author} {\bibfnamefont
  {J.}~\bibnamefont {Wrachtrup}},\ }\Doi {10.1126/science.1157233} {\bibfield
  {journal} {\bibinfo  {journal} {Science},\ }\textbf {\bibinfo {volume}
  {320}},\ \bibinfo {pages} {1326} (\bibinfo {year} {2008})}\BibitemShut
  {NoStop}%
\bibitem [{\citenamefont {Yao}\ \emph {et~al.}(2012)\citenamefont {Yao},
  \citenamefont {Wang}, \citenamefont {Xu}, \citenamefont {Lu}, \citenamefont
  {Pan}, \citenamefont {Bao}, \citenamefont {Peng}, \citenamefont {Lu},
  \citenamefont {Chen},\ and\ \citenamefont {Pan}}]{Yao2012}%
  \BibitemOpen
  \bibfield  {author} {\bibinfo {author} {\bibfnamefont {X.-C.}\ \bibnamefont
  {Yao}}, \bibinfo {author} {\bibfnamefont {T.-X.}\ \bibnamefont {Wang}},
  \bibinfo {author} {\bibfnamefont {P.}~\bibnamefont {Xu}}, \bibinfo {author}
  {\bibfnamefont {H.}~\bibnamefont {Lu}}, \bibinfo {author} {\bibfnamefont
  {G.-S.}\ \bibnamefont {Pan}}, \bibinfo {author} {\bibfnamefont {X.-H.}\
  \bibnamefont {Bao}}, \bibinfo {author} {\bibfnamefont {C.-Z.}\ \bibnamefont
  {Peng}}, \bibinfo {author} {\bibfnamefont {C.-Y.}\ \bibnamefont {Lu}},
  \bibinfo {author} {\bibfnamefont {Y.-A.}\ \bibnamefont {Chen}}, \ and\
  \bibinfo {author} {\bibfnamefont {J.-W.}\ \bibnamefont {Pan}},\ }\Doi
  {10.1038/nphoton.2011.354} {\bibfield  {journal} {\bibinfo  {journal} {Nat.
  Photon.},\ }\textbf {\bibinfo {volume} {advance online publication}}
  (\bibinfo {year} {2012})},\ \doi {10.1038/nphoton.2011.354}\BibitemShut
  {NoStop}%
\bibitem [{\citenamefont {Ursin}\ \emph {et~al.}(2007)\citenamefont {Ursin},
  \citenamefont {Tiefenbacher}, \citenamefont {Schmitt-Manderbach},
  \citenamefont {Weier}, \citenamefont {Scheidl}, \citenamefont {Lindenthal},
  \citenamefont {Blauensteiner}, \citenamefont {Jennewein}, \citenamefont
  {Perdigues}, \citenamefont {Trojek}, \citenamefont {Oemer}, \citenamefont
  {Fuerst}, \citenamefont {Meyenburg}, \citenamefont {Rarity}, \citenamefont
  {Sodnik}, \citenamefont {Barbieri}, \citenamefont {Weinfurter},\ and\
  \citenamefont {Zeilinger}}]{ISI:000248219100018}%
  \BibitemOpen
  \bibfield  {author} {\bibinfo {author} {\bibfnamefont {R.}~\bibnamefont
  {Ursin}}, \bibinfo {author} {\bibfnamefont {F.}~\bibnamefont {Tiefenbacher}},
  \bibinfo {author} {\bibfnamefont {T.}~\bibnamefont {Schmitt-Manderbach}},
  \bibinfo {author} {\bibfnamefont {H.}~\bibnamefont {Weier}}, \bibinfo
  {author} {\bibfnamefont {T.}~\bibnamefont {Scheidl}}, \bibinfo {author}
  {\bibfnamefont {M.}~\bibnamefont {Lindenthal}}, \bibinfo {author}
  {\bibfnamefont {B.}~\bibnamefont {Blauensteiner}}, \bibinfo {author}
  {\bibfnamefont {T.}~\bibnamefont {Jennewein}}, \bibinfo {author}
  {\bibfnamefont {J.}~\bibnamefont {Perdigues}}, \bibinfo {author}
  {\bibfnamefont {P.}~\bibnamefont {Trojek}}, \bibinfo {author} {\bibfnamefont
  {B.}~\bibnamefont {Oemer}}, \bibinfo {author} {\bibfnamefont
  {M.}~\bibnamefont {Fuerst}}, \bibinfo {author} {\bibfnamefont
  {M.}~\bibnamefont {Meyenburg}}, \bibinfo {author} {\bibfnamefont
  {J.}~\bibnamefont {Rarity}}, \bibinfo {author} {\bibfnamefont
  {Z.}~\bibnamefont {Sodnik}}, \bibinfo {author} {\bibfnamefont
  {C.}~\bibnamefont {Barbieri}}, \bibinfo {author} {\bibfnamefont
  {H.}~\bibnamefont {Weinfurter}}, \ and\ \bibinfo {author} {\bibfnamefont
  {A.}~\bibnamefont {Zeilinger}},\ }\Doi {10.1038/nphys629} {\bibfield
  {journal} {\bibinfo  {journal} {Nat. Phys.},\ }\textbf {\bibinfo {volume}
  {3}},\ \bibinfo {pages} {481} (\bibinfo {year} {2007})}\BibitemShut {NoStop}%
\bibitem [{\citenamefont {Lindblad}(1976)}]{lindblad}%
  \BibitemOpen
  \bibfield  {author} {\bibinfo {author} {\bibfnamefont {G.}~\bibnamefont
  {Lindblad}},\ }\href@noop {} {\bibfield  {journal} {\bibinfo  {journal}
  {Comm. Math. Phys.},\ }\textbf {\bibinfo {volume} {48}},\ \bibinfo {pages}
  {119} (\bibinfo {year} {1976})}\BibitemShut {NoStop}%
\bibitem [{\citenamefont {D'Alessandro}(2008)}]{alessandrocontrol}%
  \BibitemOpen
  \bibfield  {author} {\bibinfo {author} {\bibfnamefont {D.}~\bibnamefont
  {D'Alessandro}},\ }\href
  {file:///home/felix/Documents/Lesen/Buch/D'Alessandro - Quantum Control and
  Dynamics.pdf} {\emph {\bibinfo {title} {Introduction to {Q}uantum {C}ontrol
  and {D}ynamics}}},\ Applied Mathematics and Nonliner Science Series\
  (\bibinfo  {publisher} {Chapman \& Hall/CRC},\ \bibinfo {address} {Boca
  Raton},\ \bibinfo {year} {2008})\BibitemShut {NoStop}%
\bibitem [{\citenamefont {Platzer}\ \emph {et~al.}(2010)\citenamefont
  {Platzer}, \citenamefont {Mintert},\ and\ \citenamefont {Buchleitner}}]{PRL}%
  \BibitemOpen
  \bibfield  {author} {\bibinfo {author} {\bibfnamefont {F.}~\bibnamefont
  {Platzer}}, \bibinfo {author} {\bibfnamefont {F.}~\bibnamefont {Mintert}}, \
  and\ \bibinfo {author} {\bibfnamefont {A.}~\bibnamefont {Buchleitner}},\
  }\Doi {10.1103/PhysRevLett.105.020501} {\bibfield  {journal} {\bibinfo
  {journal} {Phys. Rev. Lett.},\ }\textbf {\bibinfo {volume} {105}},\ \bibinfo
  {pages} {020501} (\bibinfo {year} {2010})}\BibitemShut {NoStop}%
\bibitem [{\citenamefont {Tannor}\ and\ \citenamefont
  {Rice}(1985)}]{tannor:5013}%
  \BibitemOpen
  \bibfield  {author} {\bibinfo {author} {\bibfnamefont {D.~J.}\ \bibnamefont
  {Tannor}}\ and\ \bibinfo {author} {\bibfnamefont {S.~A.}\ \bibnamefont
  {Rice}},\ }\Doi {10.1063/1.449767} {\bibfield  {journal} {\bibinfo  {journal}
  {The Journal of Chemical Physics},\ }\textbf {\bibinfo {volume} {83}},\
  \bibinfo {pages} {5013} (\bibinfo {year} {1985})}\BibitemShut {NoStop}%
\bibitem [{\citenamefont {Peirce}\ \emph {et~al.}(1988)\citenamefont {Peirce},
  \citenamefont {Dahleh},\ and\ \citenamefont {Rabitz}}]{PhysRevA.37.4950}%
  \BibitemOpen
  \bibfield  {author} {\bibinfo {author} {\bibfnamefont {A.~P.}\ \bibnamefont
  {Peirce}}, \bibinfo {author} {\bibfnamefont {M.~A.}\ \bibnamefont {Dahleh}},
  \ and\ \bibinfo {author} {\bibfnamefont {H.}~\bibnamefont {Rabitz}},\ }\Doi
  {10.1103/PhysRevA.37.4950} {\bibfield  {journal} {\bibinfo  {journal} {Phys.
  Rev. A},\ }\textbf {\bibinfo {volume} {37}},\ \bibinfo {pages} {4950}
  (\bibinfo {year} {1988})}\BibitemShut {NoStop}%
\bibitem [{\citenamefont {Khaneja}\ \emph {et~al.}(2005)\citenamefont
  {Khaneja}, \citenamefont {Reiss}, \citenamefont {Kehlet}, \citenamefont
  {Schulte-Herbr\"uggen},\ and\ \citenamefont {Glaser}}]{glaserGRAPE}%
  \BibitemOpen
  \bibfield  {author} {\bibinfo {author} {\bibfnamefont {N.}~\bibnamefont
  {Khaneja}}, \bibinfo {author} {\bibfnamefont {T.}~\bibnamefont {Reiss}},
  \bibinfo {author} {\bibfnamefont {C.}~\bibnamefont {Kehlet}}, \bibinfo
  {author} {\bibfnamefont {T.}~\bibnamefont {Schulte-Herbr\"uggen}}, \ and\
  \bibinfo {author} {\bibfnamefont {S.~J.}\ \bibnamefont {Glaser}},\ }\Doi
  {DOI: 10.1016/j.jmr.2004.11.004} {\bibfield  {journal} {\bibinfo  {journal}
  {Journal of Magnetic Resonance},\ }\textbf {\bibinfo {volume} {172}},\
  \bibinfo {pages} {296} (\bibinfo {year} {2005})}\BibitemShut {NoStop}%
\bibitem [{\citenamefont {Kraus}(2010)}]{PhysRevLett.104.020504}%
  \BibitemOpen
  \bibfield  {author} {\bibinfo {author} {\bibfnamefont {B.}~\bibnamefont
  {Kraus}},\ }\Doi {10.1103/PhysRevLett.104.020504} {\bibfield  {journal}
  {\bibinfo  {journal} {Phys. Rev. Lett.},\ }\textbf {\bibinfo {volume}
  {104}},\ \bibinfo {pages} {020504} (\bibinfo {year} {2010})}\BibitemShut
  {NoStop}%
\bibitem [{\citenamefont {Carvalho}\ \emph {et~al.}(2004)\citenamefont
  {Carvalho}, \citenamefont {Mintert},\ and\ \citenamefont
  {Buchleitner}}]{flodeco}%
  \BibitemOpen
  \bibfield  {author} {\bibinfo {author} {\bibfnamefont {A.~R.~R.}\
  \bibnamefont {Carvalho}}, \bibinfo {author} {\bibfnamefont {F.}~\bibnamefont
  {Mintert}}, \ and\ \bibinfo {author} {\bibfnamefont {A.}~\bibnamefont
  {Buchleitner}},\ }\Doi {10.1103/PhysRevLett.93.230501} {\bibfield  {journal}
  {\bibinfo  {journal} {Phys. Rev. Lett.},\ }\textbf {\bibinfo {volume} {93}},\
  \bibinfo {pages} {230501} (\bibinfo {year} {2004})}\BibitemShut {NoStop}%
\bibitem [{\citenamefont {Simon}\ and\ \citenamefont
  {Kempe}(2002)}]{PhysRevA.65.052327}%
  \BibitemOpen
  \bibfield  {author} {\bibinfo {author} {\bibfnamefont {C.}~\bibnamefont
  {Simon}}\ and\ \bibinfo {author} {\bibfnamefont {J.}~\bibnamefont {Kempe}},\
  }\Doi {10.1103/PhysRevA.65.052327} {\bibfield  {journal} {\bibinfo  {journal}
  {Phys. Rev. A},\ }\textbf {\bibinfo {volume} {65}},\ \bibinfo {pages}
  {052327} (\bibinfo {year} {2002})}\BibitemShut {NoStop}%
\bibitem [{\citenamefont {Borras}\ \emph {et~al.}(2009)\citenamefont {Borras},
  \citenamefont {Majtey}, \citenamefont {Plastino}, \citenamefont {Casas},\
  and\ \citenamefont {Plastino}}]{borplasrob}%
  \BibitemOpen
  \bibfield  {author} {\bibinfo {author} {\bibfnamefont {A.}~\bibnamefont
  {Borras}}, \bibinfo {author} {\bibfnamefont {A.~P.}\ \bibnamefont {Majtey}},
  \bibinfo {author} {\bibfnamefont {A.~R.}\ \bibnamefont {Plastino}}, \bibinfo
  {author} {\bibfnamefont {M.}~\bibnamefont {Casas}}, \ and\ \bibinfo {author}
  {\bibfnamefont {A.}~\bibnamefont {Plastino}},\ }\Doi
  {10.1103/PhysRevA.79.022108} {\bibfield  {journal} {\bibinfo  {journal}
  {Phys. Rev. A},\ }\textbf {\bibinfo {volume} {79}},\ \bibinfo {eid} {022108}
  (\bibinfo {year} {2009})}\BibitemShut {NoStop}%
\bibitem [{\citenamefont {Mintert}(2010)}]{florobust}%
  \BibitemOpen
  \bibfield  {author} {\bibinfo {author} {\bibfnamefont {F.}~\bibnamefont
  {Mintert}},\ }\href@noop {} {\bibfield  {journal} {\bibinfo  {journal} {J.
  Phys. A},\ }\textbf {\bibinfo {volume} {43}},\ \bibinfo {pages} {245303}
  (\bibinfo {year} {2010})}\BibitemShut {NoStop}%
\bibitem [{\citenamefont {Wootters}(1998)}]{woocon}%
  \BibitemOpen
  \bibfield  {author} {\bibinfo {author} {\bibfnamefont {W.~K.}\ \bibnamefont
  {Wootters}},\ }\Doi {10.1103/PhysRevLett.80.2245} {\bibfield  {journal}
  {\bibinfo  {journal} {Phys. Rev. Lett.},\ }\textbf {\bibinfo {volume} {80}},\
  \bibinfo {pages} {2245} (\bibinfo {year} {1998})}\BibitemShut {NoStop}%
\bibitem [{\citenamefont {Kraus}\ and\ \citenamefont {Cirac}(2001)}]{entgates}%
  \BibitemOpen
  \bibfield  {author} {\bibinfo {author} {\bibfnamefont {B.}~\bibnamefont
  {Kraus}}\ and\ \bibinfo {author} {\bibfnamefont {J.~I.}\ \bibnamefont
  {Cirac}},\ }\Doi {10.1103/PhysRevA.63.062309} {\bibfield  {journal} {\bibinfo
   {journal} {Phys. Rev. A},\ }\textbf {\bibinfo {volume} {63}},\ \bibinfo
  {pages} {062309} (\bibinfo {year} {2001})}\BibitemShut {NoStop}%
\bibitem [{\citenamefont {D\"ur}\ \emph {et~al.}(2001)\citenamefont {D\"ur},
  \citenamefont {Vidal}, \citenamefont {Cirac}, \citenamefont {Linden},\ and\
  \citenamefont {Popescu}}]{loccon}%
  \BibitemOpen
  \bibfield  {author} {\bibinfo {author} {\bibfnamefont {W.}~\bibnamefont
  {D\"ur}}, \bibinfo {author} {\bibfnamefont {G.}~\bibnamefont {Vidal}},
  \bibinfo {author} {\bibfnamefont {J.~I.}\ \bibnamefont {Cirac}}, \bibinfo
  {author} {\bibfnamefont {N.}~\bibnamefont {Linden}}, \ and\ \bibinfo {author}
  {\bibfnamefont {S.}~\bibnamefont {Popescu}},\ }\Doi
  {10.1103/PhysRevLett.87.137901} {\bibfield  {journal} {\bibinfo  {journal}
  {Phys. Rev. Lett.},\ }\textbf {\bibinfo {volume} {87}},\ \bibinfo {pages}
  {137901} (\bibinfo {year} {2001})}\BibitemShut {NoStop}%
\bibitem [{\citenamefont {Uhlmann}(2000)}]{PhysRevA.62.032307}%
  \BibitemOpen
  \bibfield  {author} {\bibinfo {author} {\bibfnamefont {A.}~\bibnamefont
  {Uhlmann}},\ }\Doi {10.1103/PhysRevA.62.032307} {\bibfield  {journal}
  {\bibinfo  {journal} {Phys. Rev. A},\ }\textbf {\bibinfo {volume} {62}},\
  \bibinfo {pages} {032307} (\bibinfo {year} {2000})}\BibitemShut {NoStop}%
\bibitem [{\citenamefont {Mintert}\ \emph {et~al.}(2005)\citenamefont
  {Mintert}, \citenamefont {Ku\ifmmode~\acute{s}\else \'{s}\fi{}},\ and\
  \citenamefont {Buchleitner}}]{flopure}%
  \BibitemOpen
  \bibfield  {author} {\bibinfo {author} {\bibfnamefont {F.}~\bibnamefont
  {Mintert}}, \bibinfo {author} {\bibfnamefont {M.}~\bibnamefont
  {Ku\ifmmode~\acute{s}\else \'{s}\fi{}}}, \ and\ \bibinfo {author}
  {\bibfnamefont {A.}~\bibnamefont {Buchleitner}},\ }\Doi
  {10.1103/PhysRevLett.95.260502} {\bibfield  {journal} {\bibinfo  {journal}
  {Phys. Rev. Lett.},\ }\textbf {\bibinfo {volume} {95}},\ \bibinfo {pages}
  {260502} (\bibinfo {year} {2005})}\BibitemShut {NoStop}%
\bibitem [{\citenamefont {Mintert}\ and\ \citenamefont
  {Buchleitner}(2007)}]{flomix}%
  \BibitemOpen
  \bibfield  {author} {\bibinfo {author} {\bibfnamefont {F.}~\bibnamefont
  {Mintert}}\ and\ \bibinfo {author} {\bibfnamefont {A.}~\bibnamefont
  {Buchleitner}},\ }\Doi {10.1103/PhysRevLett.98.140505} {\bibfield  {journal}
  {\bibinfo  {journal} {Phys. Rev. Lett.},\ }\textbf {\bibinfo {volume} {98}},\
  \bibinfo {eid} {140505} (\bibinfo {year} {2007})}\BibitemShut {NoStop}%
\bibitem [{\citenamefont {Aolita}\ \emph {et~al.}(2008)\citenamefont {Aolita},
  \citenamefont {Buchleitner},\ and\ \citenamefont {Mintert}}]{aolita:022308}%
  \BibitemOpen
  \bibfield  {author} {\bibinfo {author} {\bibfnamefont {L.}~\bibnamefont
  {Aolita}}, \bibinfo {author} {\bibfnamefont {A.}~\bibnamefont {Buchleitner}},
  \ and\ \bibinfo {author} {\bibfnamefont {F.}~\bibnamefont {Mintert}},\ }\Doi
  {10.1103/PhysRevA.78.022308} {\bibfield  {journal} {\bibinfo  {journal}
  {Phys. Rev. A},\ }\textbf {\bibinfo {volume} {78}},\ \bibinfo {eid} {022308}
  (\bibinfo {year} {2008})}\BibitemShut {NoStop}%
\bibitem [{\citenamefont {Mirrahimi}\ \emph {et~al.}(2005)\citenamefont
  {Mirrahimi}, \citenamefont {Rouchon},\ and\ \citenamefont
  {Turinici}}]{Mirrahimi20051987}%
  \BibitemOpen
  \bibfield  {author} {\bibinfo {author} {\bibfnamefont {M.}~\bibnamefont
  {Mirrahimi}}, \bibinfo {author} {\bibfnamefont {P.}~\bibnamefont {Rouchon}},
  \ and\ \bibinfo {author} {\bibfnamefont {G.}~\bibnamefont {Turinici}},\ }\Doi
  {DOI: 10.1016/j.automatica.2005.05.018} {\bibfield  {journal} {\bibinfo
  {journal} {Automatica},\ }\textbf {\bibinfo {volume} {41}},\ \bibinfo {pages}
  {1987 } (\bibinfo {year} {2005})}\BibitemShut {NoStop}%
\bibitem [{\citenamefont {Tannor}\ \emph {et~al.}(1992)\citenamefont {Tannor},
  \citenamefont {Kazakov},\ and\ \citenamefont {Orlov}}]{tannorkrotov}%
  \BibitemOpen
  \bibfield  {author} {\bibinfo {author} {\bibfnamefont {D.}~\bibnamefont
  {Tannor}}, \bibinfo {author} {\bibfnamefont {V.}~\bibnamefont {Kazakov}}, \
  and\ \bibinfo {author} {\bibfnamefont {V.}~\bibnamefont {Orlov}},\ }\enquote
  {\bibinfo {title} {Time dependent quantum molecular dynamics},}\ \ (\bibinfo
  {publisher} {Plenum},\ \bibinfo {year} {1992})\ pp.\ \bibinfo {pages}
  {347--360}\BibitemShut {NoStop}%
\bibitem [{\citenamefont {Zhu}\ and\ \citenamefont
  {Rabitz}(1998)}]{rabitzkrotov}%
  \BibitemOpen
  \bibfield  {author} {\bibinfo {author} {\bibfnamefont {W.}~\bibnamefont
  {Zhu}}\ and\ \bibinfo {author} {\bibfnamefont {H.}~\bibnamefont {Rabitz}},\
  }\Doi {10.1063/1.476575} {\bibfield  {journal} {\bibinfo  {journal} {The
  Journal of Chemical Physics},\ }\textbf {\bibinfo {volume} {109}},\ \bibinfo
  {pages} {385} (\bibinfo {year} {1998})}\BibitemShut {NoStop}%
\bibitem [{\citenamefont {Jelezko}\ \emph {et~al.}(2004)\citenamefont
  {Jelezko}, \citenamefont {Gaebel}, \citenamefont {Popa}, \citenamefont
  {Gruber},\ and\ \citenamefont {Wrachtrup}}]{NVmanip}%
  \BibitemOpen
  \bibfield  {author} {\bibinfo {author} {\bibfnamefont {F.}~\bibnamefont
  {Jelezko}}, \bibinfo {author} {\bibfnamefont {T.}~\bibnamefont {Gaebel}},
  \bibinfo {author} {\bibfnamefont {I.}~\bibnamefont {Popa}}, \bibinfo {author}
  {\bibfnamefont {A.}~\bibnamefont {Gruber}}, \ and\ \bibinfo {author}
  {\bibfnamefont {J.}~\bibnamefont {Wrachtrup}},\ }\Doi
  {10.1103/PhysRevLett.92.076401} {\bibfield  {journal} {\bibinfo  {journal}
  {Phys. Rev. Lett.},\ }\textbf {\bibinfo {volume} {92}},\ \bibinfo {pages}
  {076401} (\bibinfo {year} {2004})}\BibitemShut {NoStop}%
\bibitem [{\citenamefont {Gershenfeld}\ and\ \citenamefont
  {Chuang}(1997)}]{NMRqc}%
  \BibitemOpen
  \bibfield  {author} {\bibinfo {author} {\bibfnamefont {N.~A.}\ \bibnamefont
  {Gershenfeld}}\ and\ \bibinfo {author} {\bibfnamefont {I.~L.}\ \bibnamefont
  {Chuang}},\ }\Doi {10.1126/science.275.5298.350} {\bibfield  {journal}
  {\bibinfo  {journal} {Science},\ }\textbf {\bibinfo {volume} {275}},\
  \bibinfo {pages} {350} (\bibinfo {year} {1997})}\BibitemShut {NoStop}%
\bibitem [{\citenamefont {Gaebel}\ \emph {et~al.}(2006)\citenamefont {Gaebel},
  \citenamefont {Domhan}, \citenamefont {Popa}, \citenamefont {Wittmann},
  \citenamefont {Neumann}, \citenamefont {Jelezko}, \citenamefont {Rabeau},
  \citenamefont {Stavrias}, \citenamefont {Greentree}, \citenamefont {Prawer},
  \citenamefont {Meijer}, \citenamefont {Twamley}, \citenamefont {Hemmer},\
  and\ \citenamefont {Wrachtrup}}]{Gaebel2006408}%
  \BibitemOpen
  \bibfield  {author} {\bibinfo {author} {\bibfnamefont {T.}~\bibnamefont
  {Gaebel}}, \bibinfo {author} {\bibfnamefont {M.}~\bibnamefont {Domhan}},
  \bibinfo {author} {\bibfnamefont {I.}~\bibnamefont {Popa}}, \bibinfo {author}
  {\bibfnamefont {C.}~\bibnamefont {Wittmann}}, \bibinfo {author}
  {\bibfnamefont {P.}~\bibnamefont {Neumann}}, \bibinfo {author} {\bibfnamefont
  {F.}~\bibnamefont {Jelezko}}, \bibinfo {author} {\bibfnamefont {J.~R.}\
  \bibnamefont {Rabeau}}, \bibinfo {author} {\bibfnamefont {N.}~\bibnamefont
  {Stavrias}}, \bibinfo {author} {\bibfnamefont {A.~D.}\ \bibnamefont
  {Greentree}}, \bibinfo {author} {\bibfnamefont {S.}~\bibnamefont {Prawer}},
  \bibinfo {author} {\bibfnamefont {J.}~\bibnamefont {Meijer}}, \bibinfo
  {author} {\bibfnamefont {J.}~\bibnamefont {Twamley}}, \bibinfo {author}
  {\bibfnamefont {P.~R.}\ \bibnamefont {Hemmer}}, \ and\ \bibinfo {author}
  {\bibfnamefont {J.}~\bibnamefont {Wrachtrup}},\ }\href
  {file:///home/felix/Documents/Article_PRA/references/nphys318.pdf} {\bibfield
   {journal} {\bibinfo  {journal} {Nature Physics},\ }\textbf {\bibinfo
  {volume} {2}},\ \bibinfo {pages} {408} (\bibinfo {year} {2006})}\BibitemShut
  {NoStop}%
\bibitem [{\citenamefont {Balasubramanian}\ \emph {et~al.}(2009)\citenamefont
  {Balasubramanian}, \citenamefont {Neumann}, \citenamefont {Twitchen},
  \citenamefont {Markham}, \citenamefont {Kolesov}, \citenamefont {Mizuochi},
  \citenamefont {Isoya}, \citenamefont {Achard}, \citenamefont {Beck},
  \citenamefont {Tissler}, \citenamefont {Jacques}, \citenamefont {Hemmer},
  \citenamefont {Jelezko},\ and\ \citenamefont {Wrachtrup}}]{NVcohtime}%
  \BibitemOpen
  \bibfield  {author} {\bibinfo {author} {\bibfnamefont {G.}~\bibnamefont
  {Balasubramanian}}, \bibinfo {author} {\bibfnamefont {P.}~\bibnamefont
  {Neumann}}, \bibinfo {author} {\bibfnamefont {D.}~\bibnamefont {Twitchen}},
  \bibinfo {author} {\bibfnamefont {M.}~\bibnamefont {Markham}}, \bibinfo
  {author} {\bibfnamefont {R.}~\bibnamefont {Kolesov}}, \bibinfo {author}
  {\bibfnamefont {N.}~\bibnamefont {Mizuochi}}, \bibinfo {author}
  {\bibfnamefont {J.}~\bibnamefont {Isoya}}, \bibinfo {author} {\bibfnamefont
  {J.}~\bibnamefont {Achard}}, \bibinfo {author} {\bibfnamefont
  {J.}~\bibnamefont {Beck}}, \bibinfo {author} {\bibfnamefont {J.}~\bibnamefont
  {Tissler}}, \bibinfo {author} {\bibfnamefont {V.}~\bibnamefont {Jacques}},
  \bibinfo {author} {\bibfnamefont {P.~R.}\ \bibnamefont {Hemmer}}, \bibinfo
  {author} {\bibfnamefont {F.}~\bibnamefont {Jelezko}}, \ and\ \bibinfo
  {author} {\bibfnamefont {J.}~\bibnamefont {Wrachtrup}},\ }\Doi
  {10.1038/nmat2420} {\bibfield  {journal} {\bibinfo  {journal} {Nature
  Materials},\ }\textbf {\bibinfo {volume} {8}},\ \bibinfo {pages} {383}
  (\bibinfo {year} {2009})}\BibitemShut {NoStop}%
\bibitem [{Note1()}]{Note1}%
  \BibitemOpen
  \bibinfo {note} {This can be seen by analytically solving Eq.~\protect
  \textup {\hbox {\mathsurround \z@ \protect \normalfont (\ignorespaces \ref
  {eq:drho}\unskip \@@italiccorr )}} for $L_i = \sigma _z^{(i)}$ and
  $H_\protect \text {c} = H_\protect \text {sys}=0$.}\BibitemShut {Stop}%
\bibitem [{\citenamefont {D\"ur}\ and\ \citenamefont
  {Briegel}(2004)}]{PhysRevLett.92.180403}%
  \BibitemOpen
  \bibfield  {author} {\bibinfo {author} {\bibfnamefont {W.}~\bibnamefont
  {D\"ur}}\ and\ \bibinfo {author} {\bibfnamefont {H.-J.}\ \bibnamefont
  {Briegel}},\ }\Doi {10.1103/PhysRevLett.92.180403} {\bibfield  {journal}
  {\bibinfo  {journal} {Phys. Rev. Lett.},\ }\textbf {\bibinfo {volume} {92}},\
  \bibinfo {pages} {180403} (\bibinfo {year} {2004})}\BibitemShut {NoStop}%
\bibitem [{\citenamefont {Haar}(1933)}]{haarmeasure}%
  \BibitemOpen
  \bibfield  {author} {\bibinfo {author} {\bibfnamefont {A.}~\bibnamefont
  {Haar}},\ }\href {http://www.jstor.org/stable/1968346} {\bibfield  {journal}
  {\bibinfo  {journal} {Ann. Math.},\ }\bibinfo {series} {2},\ \textbf
  {\bibinfo {volume} {34}},\ \bibinfo {pages} {pp. 147} (\bibinfo {year}
  {1933})}\BibitemShut {NoStop}%
\bibitem [{\citenamefont {Sarovar}\ \emph {et~al.}(2010)\citenamefont
  {Sarovar}, \citenamefont {Ishizaki}, \citenamefont {Fleming},\ and\
  \citenamefont {Whaley}}]{ISI:000279014400020}%
  \BibitemOpen
  \bibfield  {author} {\bibinfo {author} {\bibfnamefont {M.}~\bibnamefont
  {Sarovar}}, \bibinfo {author} {\bibfnamefont {A.}~\bibnamefont {Ishizaki}},
  \bibinfo {author} {\bibfnamefont {G.~R.}\ \bibnamefont {Fleming}}, \ and\
  \bibinfo {author} {\bibfnamefont {K.~B.}\ \bibnamefont {Whaley}},\ }\Doi
  {10.1038/NPHYS1652} {\bibfield  {journal} {\bibinfo  {journal} {Nat. Phys.},\
  }\textbf {\bibinfo {volume} {6}},\ \bibinfo {pages} {462} (\bibinfo {year}
  {2010})}\BibitemShut {NoStop}%
\bibitem [{\citenamefont {{Tiersch}}\ \emph {et~al.}(2011)\citenamefont
  {{Tiersch}}, \citenamefont {{Popescu}},\ and\ \citenamefont
  {{Briegel}}}]{2011arXiv1104.3883T}%
  \BibitemOpen
  \bibfield  {author} {\bibinfo {author} {\bibfnamefont {M.}~\bibnamefont
  {{Tiersch}}}, \bibinfo {author} {\bibfnamefont {S.}~\bibnamefont
  {{Popescu}}}, \ and\ \bibinfo {author} {\bibfnamefont {H.~J.}\ \bibnamefont
  {{Briegel}}},\ }\href@noop {} {\bibfield  {journal} {\bibinfo  {journal}
  {ArXiv e-prints}} (\bibinfo {year} {2011})},\ \Eprint
  {http://arxiv.org/abs/1104.3883} {arXiv:1104.3883 [quant-ph]} \BibitemShut
  {NoStop}%
\bibitem [{\citenamefont {Gross}\ \emph {et~al.}(2009)\citenamefont {Gross},
  \citenamefont {Flammia},\ and\ \citenamefont {Eisert}}]{gross:190501}%
  \BibitemOpen
  \bibfield  {author} {\bibinfo {author} {\bibfnamefont {D.}~\bibnamefont
  {Gross}}, \bibinfo {author} {\bibfnamefont {S.~T.}\ \bibnamefont {Flammia}},
  \ and\ \bibinfo {author} {\bibfnamefont {J.}~\bibnamefont {Eisert}},\ }\Doi
  {10.1103/PhysRevLett.102.190501} {\bibfield  {journal} {\bibinfo  {journal}
  {Phys. Rev. Lett.},\ }\textbf {\bibinfo {volume} {102}},\ \bibinfo {eid}
  {190501--190504} (\bibinfo {year} {2009})}\BibitemShut {NoStop}%
\bibitem [{\citenamefont {M\"uller}\ \emph {et~al.}(2011)\citenamefont
  {M\"uller}, \citenamefont {Reich}, \citenamefont {Murphy}, \citenamefont
  {Yuan}, \citenamefont {Vala}, \citenamefont {Whaley}, \citenamefont
  {Calarco},\ and\ \citenamefont {Koch}}]{KochEntGate}%
  \BibitemOpen
  \bibfield  {author} {\bibinfo {author} {\bibfnamefont {M.~M.}\ \bibnamefont
  {M\"uller}}, \bibinfo {author} {\bibfnamefont {D.~M.}\ \bibnamefont {Reich}},
  \bibinfo {author} {\bibfnamefont {M.}~\bibnamefont {Murphy}}, \bibinfo
  {author} {\bibfnamefont {H.}~\bibnamefont {Yuan}}, \bibinfo {author}
  {\bibfnamefont {J.}~\bibnamefont {Vala}}, \bibinfo {author} {\bibfnamefont
  {K.~B.}\ \bibnamefont {Whaley}}, \bibinfo {author} {\bibfnamefont
  {T.}~\bibnamefont {Calarco}}, \ and\ \bibinfo {author} {\bibfnamefont
  {C.~P.}\ \bibnamefont {Koch}},\ }\Doi {10.1103/PhysRevA.84.042315} {\bibfield
   {journal} {\bibinfo  {journal} {Phys. Rev. A},\ }\textbf {\bibinfo {volume}
  {84}},\ \bibinfo {pages} {042315} (\bibinfo {year} {2011})}\BibitemShut
  {NoStop}%
\bibitem [{\citenamefont {Mintert}\ \emph {et~al.}(2011)\citenamefont
  {Mintert}, \citenamefont {Lapert}, \citenamefont {Zhang}, \citenamefont
  {Glaser},\ and\ \citenamefont {Sugny}}]{floglaser}%
  \BibitemOpen
  \bibfield  {author} {\bibinfo {author} {\bibfnamefont {F.}~\bibnamefont
  {Mintert}}, \bibinfo {author} {\bibfnamefont {M.}~\bibnamefont {Lapert}},
  \bibinfo {author} {\bibfnamefont {Y.}~\bibnamefont {Zhang}}, \bibinfo
  {author} {\bibfnamefont {S.~J.}\ \bibnamefont {Glaser}}, \ and\ \bibinfo
  {author} {\bibfnamefont {D.}~\bibnamefont {Sugny}},\ }\href
  {http://stacks.iop.org/1367-2630/13/i=7/a=073001} {\bibfield  {journal}
  {\bibinfo  {journal} {New Journal of Physics},\ }\textbf {\bibinfo {volume}
  {13}},\ \bibinfo {pages} {073001} (\bibinfo {year} {2011})}\BibitemShut
  {NoStop}%
\bibitem [{\citenamefont {Engel}\ \emph {et~al.}(2007)\citenamefont {Engel},
  \citenamefont {Calhoun}, \citenamefont {Read}, \citenamefont {Ahn},
  \citenamefont {Mancal}, \citenamefont {Cheng}, \citenamefont {Blankenship},\
  and\ \citenamefont {Fleming}}]{engel}%
  \BibitemOpen
  \bibfield  {author} {\bibinfo {author} {\bibfnamefont {G.}~\bibnamefont
  {Engel}}, \bibinfo {author} {\bibfnamefont {T.}~\bibnamefont {Calhoun}},
  \bibinfo {author} {\bibfnamefont {E.}~\bibnamefont {Read}}, \bibinfo {author}
  {\bibfnamefont {T.-K.}\ \bibnamefont {Ahn}}, \bibinfo {author} {\bibfnamefont
  {T.}~\bibnamefont {Mancal}}, \bibinfo {author} {\bibfnamefont {Y.-C.}\
  \bibnamefont {Cheng}}, \bibinfo {author} {\bibfnamefont {R.}~\bibnamefont
  {Blankenship}}, \ and\ \bibinfo {author} {\bibfnamefont {G.}~\bibnamefont
  {Fleming}},\ }\href@noop {} {\bibfield  {journal} {\bibinfo  {journal}
  {Nature},\ }\textbf {\bibinfo {volume} {446}},\ \bibinfo {pages} {782}
  (\bibinfo {year} {2007})}\BibitemShut {NoStop}%
\bibitem [{\citenamefont {Scholes}\ \emph {et~al.}(2011)\citenamefont
  {Scholes}, \citenamefont {Fleming}, \citenamefont {Olaya-Castro},\ and\
  \citenamefont {van Grondelle}}]{Scholes2011}%
  \BibitemOpen
  \bibfield  {author} {\bibinfo {author} {\bibfnamefont {G.~D.}\ \bibnamefont
  {Scholes}}, \bibinfo {author} {\bibfnamefont {G.~R.}\ \bibnamefont
  {Fleming}}, \bibinfo {author} {\bibfnamefont {A.}~\bibnamefont
  {Olaya-Castro}}, \ and\ \bibinfo {author} {\bibfnamefont {R.}~\bibnamefont
  {van Grondelle}},\ }\Doi {10.1038/nchem.1145} {\bibfield  {journal} {\bibinfo
   {journal} {Nat. Chem.},\ }\textbf {\bibinfo {volume} {3}},\ \bibinfo {pages}
  {763} (\bibinfo {year} {2011})}\BibitemShut {NoStop}%
\bibitem [{\citenamefont {Zewail}\ and\ \citenamefont
  {Bernstein}(1988)}]{zewailbondbreak}%
  \BibitemOpen
  \bibfield  {author} {\bibinfo {author} {\bibfnamefont {A.}~\bibnamefont
  {Zewail}}\ and\ \bibinfo {author} {\bibfnamefont {R.}~\bibnamefont
  {Bernstein}},\ }\href@noop {} {\bibfield  {journal} {\bibinfo  {journal}
  {Chem. Eng. News},\ }\textbf {\bibinfo {volume} {66}},\ \bibinfo {pages} {24}
  (\bibinfo {year} {1988})}\BibitemShut {NoStop}%
\end{thebibliography}%

\end{document}